\documentclass[12pt]{article}
\usepackage{epsf}
\def\baselinestretch{1.2}
\makeatletter
\@addtoreset{equation}{section}
\renewcommand{\theequation}{\thesection.\arabic{equation}}
\renewcommand{\thefootnote}{\fnsymbol{footnote}}
\makeatother

\begin{document}
\thispagestyle{empty}
%\begin{titlepage}
%
\begin{flushright}
TIT/HEP--487 \\
{\tt hep-th/0211103} \\
November, 2002 \\
\end{flushright}
\vspace{3mm}
\begin{center}
{\Large
{\bf Manifest Supersymmetry for BPS Walls 
%Effective Action on the Wall in Supersymmetric Theories
} 
\\
\vspace{2mm}{\bf 
in ${\cal N}=2$ Nonlinear Sigma Models
}
} 
\\[12mm]
\vspace{5mm}

\normalsize
  {\large \bf 
  Masato~Arai~$^{a}$}
\footnote{Present address is {\it Institute of Physics, AS CR, 
  182 21, Praha 8, Czech Republic}.
}
\footnote{\it  e-mail address: 
arai@fzu.cz}
,  
  {\large \bf 
  Masashi~Naganuma~$^{a}$}
\footnote{\it  e-mail address: 
naganuma@th.phys.titech.ac.jp
},  
 {\large \bf 
Muneto~Nitta~$^{b}$}
\footnote{\it  e-mail address: 
nitta@physics.purdue.edu
},
~and~~  {\large \bf 
Norisuke~Sakai~$^{a}$}
\footnote{\it  e-mail address: 
nsakai@th.phys.titech.ac.jp
} 

\vskip 1.5em

{ \it $^{a}$Department of Physics, Tokyo Institute of 
Technology \\
Tokyo 152-8551, JAPAN  \\
and \\
  $^{b}$
  Department of Physics, Purdue University, 
West Lafayette, IN 47907-1396, USA
 }
\vspace{5mm}
{\bf Abstract}\\[5mm]
{\parbox{13cm}{\hspace{5mm}
%%%%%%%%%%%%%%%%%%%%%%%%%%%%%%%%%%%%%%%%%%%%%%%%%%
%%%%%%%%%%
BPS equations and %BPS 
wall solutions are studied 
keeping (part of) supersymmetry (SUSY) manifest. 
Using ${\cal N}=1$ superfields, 
 massive hyper-K\"ahler quotient is introduced to obtain 
 massive ${\cal N}=2$ 
 ($8$ SUSY) nonlinear sigma models in four dimensions with 
% the cotangent bundle over ${\bf C}P^n$, $T^*{\bf C}P^n$ 
 $T^*{\bf C}P^n$ 
 target manifold, which yield 
 BPS wall solutions for the $n=1$ case. 
We also describe massive hyper-K\"ahler quotient of 
 $T^*{\bf C}P^n$ by using the harmonic superspace
 formalism 
 which preserves all SUSY manifestly, and
 BPS equations and wall solutions are obtained 
 in the $n=1$ case.
%%%%%%%%%%%%%%%%%%%%%%%%%%%%%%%%%%%%%%%%%%%%%%%%%%
%%%%%%%%%%
}}
\end{center}
\vfill
\newpage
\setcounter{page}{1}
\setcounter{footnote}{0}
\renewcommand{\thefootnote}{\arabic{footnote}}

%%%%%%%%%%%%%%%%%%%%%%%%%%%%%%%%%%%%%%
\section{Introduction}\label{INTRO}
%%%%%%%%%%%%%%%%%%%%%%%%%%%%%%%%%%%%%%
\vspace{5mm}

Brane world scenario has attracted much attention in recent 
 years \cite{LED,RS}. 
To realize our world on a brane such as a wall, it is 
 useful to consider supersymmetric (SUSY) theories which is 
 the most promising theory beyond the standard model \cite{DGSW}. 
Wall (junction) configurations can preserve half (quarter) 
 of the SUSY charges \cite{CGR}--\cite{NNS2} and are called 
 $\frac{1}{2}$ ($\frac{1}{4}$) BPS states.
To obtain four dimensional brane as a wall/or junction, we have to
 consider a SUSY fundamental theory in five or more spacetime
 dimensions, and such a theory must have at least eight SUSY 
 charges \cite{WB}.
These SUSY are so restrictive that possible potential terms
 are severely constrained.
The nontrivial interactions require either nonlinearity 
 of kinetic term (nonlinear sigma model) or gauge interactions
 \cite{SierraTownsend}--\cite{NNS1}.
If the theory is dimensionally reduced to four dimensions, 
 it has at least ${\cal N}=2$ SUSY.
In the ${\cal N}=2$ SUSY in four dimensions, one has to consider
 nonlinear sigma model with nontrivial K{\"a}hler metric 
 in field space
 if one wants interacting theories 
%to have nontrivial interaction terms 
with only hypermultiplets.
The target spaces of such theories
 must be hyper-K{\"a}hler (HK) manifolds~\cite{Zu}. 
%Only Possible p
Potential terms 
%are 
can be induced only when masses of
 hypermultiplets are introduced.
It was shown that its form can be described by the norm of 
 Killing vector of target metric \cite{AF,GTT1}.
Therefore, we have to consider massive nonlinear sigma models for
 ${\cal N}=2$ SUSY theories if one wants to obtain
 an interesting solution like domain walls using only the hypermultiplets.

There have been a number of works to study the nonlinear sigma models 
 with eight supercharges \cite{SierraTownsend}--\cite{LR}. 
The massive nonlinear sigma
 model with 
 nontrivial K{\"a}hler metric as target space was
 studied, and BPS equations and
 BPS wall/or junction solutions were obtained \cite{AT}--\cite{PT}. 
Multi domain walls solution was also 
 obtained
 and the dynamics of those walls was examined \cite{GTT,To}.
In most papers, nonlinear sigma models were studied in
terms of component fields.
However, it is often useful to maintain as much SUSY as possible. 
For instance, 
 maintaining three-dimensional SUSY is 
 useful to obtain a low energy effective action (LEEA) on the 
 wall in four-dimensional SUSY as fundamental 
 theory \cite{Sakamura}. 
Harmonic superspace 
 formalism (HSF) \cite{Ivanov} is most suited to maintain the SUSY 
 maximally, but there has been relatively few attempt to
 formulate the BPS equations and to obtain BPS 
 solutions in the HSF \cite{Zupnik}.

The purpose of our paper is to present SUSY 
 formulation of BPS equations and solutions in 
 ${\cal N}=2$ nonlinear sigma models in four dimensions 
 using both ${\cal N}=1$ superfield 
 and ${\cal N}=2$ superfield (HSF). 
Action of nonlinear sigma model can be constructed 
by the method of HK quotient in both languages. 
Furthermore, one has to introduce masses for hypermultiplets
 in order to obtain a scalar potential with nontrivial
 interactions.
Therefore, we need to construct a HK quotient
 for the massive nonlinear sigma model. 
We call such a quotient massive HK quotient.
In this paper, we construct massive HK 
 quotient of nonlinear sigma model on 
the cotangent bundle over ${\bf C}P^n$, namely 
$T^*{\bf C}P^{n}$, 
 with the potential term
 and describe the action in both ${\cal N}=1$ and ${\cal N}=2$ 
 formalisms.
In ${\cal N}=1$ formalism,
 the massless HK sigma model on $T^*{\bf C}P^{n}$ 
 was obtained as the HK quotient~\cite{RT,LR}.
The massive HK quotient 
 was obtained in component level~\cite{To}. 
In ${\cal N}=2$ formalism,
 the massless model on $T^*{\bf C}P^{1}$ was first  
 constructed in Ref.~\cite{ivanov-EH}, and
 its central extension was analysed in Ref.~\cite{Ketov}.
The massive HK sigma model on $T^*{\bf C}P^{n}$ 
 in superfield languages can be easily
 obtained as an extension of above models.
Since $T^*{\bf C}P^n$ is one of the simplest classes of HK 
 manifolds, we anticipate these quotient constructions to be useful 
 in future. 

In this paper, as the simplest case,
 we focus on the $n=1$ case, massive $T^*{\bf C}P^1$
 model and examine the BPS equations and wall solutions.
We show that the solution which is obtained from 
 both languages corresponds to that derived in 
the component formalism \cite{GTT1}.
Our formalism hopefully provides a starting point to obtain solutions 
and effective actions in more realistic 
cases of four-dimensional wall in five or higher dimensional theories.

Four is the minimum number of SUSY charges in four dimensions. 
Hence one might expect that 
 one could deal with the models in the manner 
 where four SUSY conserved by BPS wall is manifest, 
 since BPS walls in ${\cal N}=2$ nonlinear sigma models 
 conserve four SUSY. 
However, we cannot construct Lorentz invariant field theory keeping 
 only those four SUSY conserved by the wall manifestly, 
 because half 
 SUSY condition for BPS wall itself breaks the Lorentz invariance for four 
 dimensional spacetime.
 For instance, 
the Weyl spinor parameters of ${\cal N}=2$ SUSY $\epsilon^i~(i=1,2)$ 
are constrained by the ${1 \over 2}$ SUSY condition for 
BPS wall solutions depending on 
 a single coordinate $y$  
 (with a simple choice of spinor basis and of parameters) 
%%%%%%%%%%%%%%%%%%%%%%%%%%%%%
\begin{eqnarray}
  \sigma^2 \bar{\epsilon}^1 = i \epsilon^1, \quad 
  \sigma^2 \bar{\epsilon}^2 = -i \epsilon^2 \label{eq:half-SUSY}
\end{eqnarray} 
%%%%%%%%%%%%%%%%%%%%%%%%%%%%%
as derived 
\footnote{
We follow mostly the notation of Ref.~\cite{WB}, 
except that $\mu, \nu, \dots$ denote space time in four dimensions, 
$a, b \dots$ three dimensions on the wall. 
}
in Appendix \ref{half-SUSY}. 
The four SUSY selected by this condition 
 allows the model to be Lorentz invariant 
 for only three dimensional spacetime $(t,x,z)$ corresponding 
 to the world volume 
 of the wall. 

In sect.~2, we introduce HK quotient methods 
 in the ${\cal N}=1$ superfield formalism to construct 
 massive nonlinear sigma models with 
 $T^*{\bf C}P^n$ target space manifolds. 
We examine BPS equations and a wall solution in the $n=1$ case.
In sect.~3, we study the same contents in the previous section by 
 using the HSF. 
In sect.~4,  the relation between the usual method to obtain 
% the low energy effective action (
LEEA %)
 of zero modes 
 and the Manton's approach \cite{Manton} is discussed. 
%Sect.~4 is devoted to discuss the low energy effective action 
%(LEEA) of massive $T^*{\bf C}P^1$ model.
In Appendices we explain some details of 
 Lorentz invariance versus $\frac{1}{2}$ BPS condition, HSF, and
 some topics related to each section.

%%%%%%%%%%%%%%%%%%%%%%
%                    %
% Section 2 MHKQ     %
%                    %
%%%%%%%%%%%%%%%%%%%%%%
%\include{mhkq4}
%%%%%%%%%%%%%%%%%%%%%%%%%%%%%%%%%%%%%%
% Massive HK quotient
%%%%%%%%%%%%%%%%%%%%%%%%%%%%%%%%%%%%%%
%
% 8/11/2002 Nitta
%
% the version with b
% mhkq4.tex
%
%\documentstyle[12pt,showkeys]{article}
%\documentstyle[12pt]{article}

%\renewcommand{\baselinestretch}{1.4}
%\textwidth 150mm
%\textheight 220mm

%\makeatletter
%\renewcommand{\theequation}{%
%   \thesection.\arabic{equation}}
% \@addtoreset{equation}{section}
%\makeatother

%\begin{document}
%\topmargin 0pt
%\oddsidemargin 5mm

%%%%%%  user's commands  %%%%%%%%%%%%%%%%%%%%%%%%%%%%%%%%%%%%%%%%%%%
\newcommand {\beq}{\begin{eqnarray}}
\newcommand {\eeq}{\end{eqnarray}}
\newcommand {\non}{\nonumber\\}
\newcommand {\eq}[1]{\label {eq.#1}}
\newcommand {\defeq}{\stackrel{\rm def}{=}}
\newcommand {\gto}{\stackrel{g}{\to}}
\newcommand {\hto}{\stackrel{h}{\to}}
\newcommand {\1}[1]{\frac{1}{#1}}
\newcommand {\2}[1]{\frac{i}{#1}}
\newcommand {\thb}{\bar{\theta}}
\newcommand {\ps}{\psi}
\newcommand {\psb}{\bar{\psi}}
\newcommand {\ph}{\varphi}
\newcommand {\phs}[1]{\varphi^{*#1}}
\newcommand {\sig}{\sigma}
\newcommand {\sigb}{\bar{\sigma}}
\newcommand {\Ph}{\Phi}
\newcommand {\Phd}{\Phi^{\dagger}}
\newcommand {\Sig}{\Sigma}
\newcommand {\Phm}{{\mit\Phi}}
\newcommand {\eps}{\varepsilon}
\newcommand {\del}{\partial}
\newcommand {\dagg}{^{\dagger}}
\newcommand {\pri}{^{\prime}}
\newcommand {\prip}{^{\prime\prime}}
\newcommand {\pripp}{^{\prime\prime\prime}}
\newcommand {\prippp}{^{\prime\prime\prime\prime}}
\newcommand {\pripppp}{^{\prime\prime\prime\prime\prime}}
\newcommand {\delb}{\bar{\partial}}
\newcommand {\zb}{\bar{z}}
\newcommand {\mub}{\bar{\mu}}
\newcommand {\nub}{\bar{\nu}}
\newcommand {\lam}{\lambda}
\newcommand {\lamb}{\bar{\lambda}}
\newcommand {\kap}{\kappa}
\newcommand {\kapb}{\bar{\kappa}}
\newcommand {\xib}{\bar{\xi}}
\newcommand {\ep}{\epsilon}
\newcommand {\epb}{\bar{\epsilon}}
\newcommand {\Ga}{\Gamma}
\newcommand {\rhob}{\bar{\rho}}
\newcommand {\etab}{\bar{\eta}}
\newcommand {\chib}{\bar{\chi}}
\newcommand {\tht}{\tilde{\th}}
\newcommand {\zbasis}[1]{\del/\del z^{#1}}
\newcommand {\zbbasis}[1]{\del/\del \bar{z}^{#1}}
\newcommand {\vecv}{\vec{v}^{\, \prime}}
\newcommand {\vecvd}{\vec{v}^{\, \prime \dagger}}
\newcommand {\vecvs}{\vec{v}^{\, \prime *}}
\newcommand {\alpht}{\tilde{\alpha}}
\newcommand {\xipd}{\xi^{\prime\dagger}}
\newcommand {\pris}{^{\prime *}}
\newcommand {\prid}{^{\prime \dagger}}
\newcommand {\Jto}{\stackrel{J}{\to}}
\newcommand {\vprid}{v^{\prime 2}}
\newcommand {\vpriq}{v^{\prime 4}}
\newcommand {\vt}{\tilde{v}}
\newcommand {\vecvt}{\vec{\tilde{v}}}
\newcommand {\vecpht}{\vec{\tilde{\phi}}}
\newcommand {\pht}{\tilde{\phi}}
\newcommand {\goto}{\stackrel{g_0}{\to}}
\newcommand {\tr}{{\rm tr}\,}
\newcommand {\GC}{G^{\bf C}}
\newcommand {\HC}{H^{\bf C}}
\newcommand{\vs}[1]{\vspace{#1 mm}}
\newcommand{\hs}[1]{\hspace{#1 mm}}
\newcommand{\al}{\alpha}
\newcommand{\be}{\beta}
\newcommand{\Lam}{\Lambda}

\newcommand{\kahler}{K\"ahler }
\newcommand{\con}[1]{{\Gamma^{#1}}}

%%%%%%%%%%%%%%%%%%%%%%%%%%%%%%%%%%%%%%%%%%%%%%%%%%%%%%%%%%
\section{Massive hyper-K\"ahler quotient 
with ${\cal N}=1$ superfield 
}\label{sc:MHKQ}
In this section, 
 the massive HK sigma model on 
% the cotangent bundle over ${\bf C}P^{n}$, namely 
$T^*{\bf C}P^{n}$
 is obtained in the ${\cal N}=1$ superfield formalism.  
We show that the potential term is the square of 
 a tri-holomorphic Killing vector on the manifold 
 as shown in the component level by Ref.~\cite{GTT1}. 
We then obtain the BPS wall solution in the massive HK sigma model 
 on $T^*{\bf C}P^1$. 
In appendix B, we give another ${\cal N}=1$ superfield formulation 
 of the massive $T^*{\bf C}P^1$ model and its BPS wall solution, 
 with generalisation 
 to the Gibbons-Hawking metric \cite{GH} 
 using the Hitchin method \cite{Hi}.

%%%%%
\subsection{Massive HK sigma model on $T^* {\bf C}P^{n}$}
In this subsection,
 we construct massive HK sigma model on $T^*{\bf C}P^{n}$ 
 by using the HK quotient in ${\cal N}=1$ superfield 
 formalism.

An ${\cal N}=2$ hypermultiplet can be decomposed into 
 two chiral superfields in the ${\cal N}=1$ superfield formalism. 
We decompose ($n+1$)-hypermultiplets belonging to 
 the fundamental representation of $SU(n+1)$ into 
 ${\cal N}=1$ chiral superfields 
 $\phi(x,\theta,\thb) = (\phi^1 ,\cdots, \phi^{n+1})^T$ 
 and $\chi(x,\theta,\thb) = (\chi^1 ,\cdots, \chi^{n+1})^T$,  
 belonging to the fundamental and anti-fundamental representations  
 of $SU(n+1)$, respectively, 
 whose transformation laws under $SU(n+1)$ are given by 
\beq
 \phi \to \phi' = g \phi \;, \hs{5}
 \chi \to \chi' = (g^{-1})^T \chi \;,  \label{global}
\eeq
 with $g \in SU(n+1)$. 
An ${\cal N}=2$ vector superfield of 
 the $U(1)$ gauge symmetry can be 
 decomposed into 
 ${\cal N}=1$ vector and chiral superfields, 
 $V(x,\theta,\thb)$ and $\sig(x,\theta,\thb)$.  
The $U(1)$-charges of $\phi$ and $\chi$ are $1$ and $-1$, respectively.   
The $U(1)$ gauge transformation is given by 
\beq
 e^V \to e^{V'} = e^{i \Lam - i \Lam\dagg} e^V  \;, \hs{5}
 \phi \to \phi' = e^{i \Lam} \phi \;, \hs{5} 
 \chi \to \chi' = e^{-i \Lam} \chi \;, \label{gauge-sym}
\eeq
where $\Lam(x,\theta,\theta)$ is a chiral superfield of 
 a gauge parameter. Note that this $U(1)$ gauge symmetry 
 is actually enhanced to its complexification, 
 $U(1)^{\bf C}$. 
Then the Lagrangian of the $(n+1)$-hypermultiplets interacting 
 with the auxiliary vector multiplet can be given by\footnote{
We omitted the trace part of the mass term $\chi^T\cdot\phi$, 
 since it vanishes under the superspace integration 
 using the constraint (\ref{EOM-sig}), in the following.
} 
\beq
 {\cal L} &=& \int d^4 \theta 
 (e^V \phi\dagg\phi + e^{-V} \chi\dagg\chi - c V) 
 + \left( \int d^2 \theta \sig (\chi^T \cdot \phi - b) 
         + {\rm c.c.}\right)
 \non 
 && + \left( \int d^2 \theta \sum_{{\alpha}=1}^{n} m_{\alpha} \chi^T 
 H_{\alpha} \phi 
 + {\rm c.c.}\right) \;, \label{linear}
\eeq
 where the dot denotes the inner product of two vectors,  
 $H_{\alpha}$ (${\alpha}=1,\cdots ,n$) 
 are the diagonal generators of the Cartan subalgebra of $SU(n+1)$ 
 and $m_{\alpha}$ are the $n$ mass parameters. 
Here $ c \in {\bf R} $ and $ b \in {\bf C} $ are 
 coefficients of the Fayet-Iliopoulos (FI) term which 
 transforms as an $ SU(2)_R $ triplet
(See Lagrangian (\ref{TSCPN}) in the next section.) 
\cite{Fayet}.

In the limit of $m_{\alpha} =0$ for all ${\alpha}$, 
 the Lagrangian (\ref{linear}) becomes that of 
 the massless HK 
 nonlinear sigma model on $T^* {\bf C}P^{n}$~\cite{CF}--\cite{LR}  
 whose isometry is $SU(n+1)$. 
By introducing the mass $m_\alpha \not=0$ of the last term 
in (\ref{linear}), we will obtain the massive HK nonlinear 
 sigma model on $T^* {\bf C}P^{n}$. 
The mass term explicitly breaks $SU(n+1)$ into $U(1)^{n}$: 
 if we write $g= e^{i\eps^A T_A}$ in (\ref{global})
 where $T_A$ are the fundamental representation of 
 the generators of $SU(n+1)$ ($A=1, \cdots (n+1)^2-1$), 
the infinitesimal variation of the mass term 
\beq
 \delta_{\eps} 
 (\sum_{{\alpha}=1}^{n} m_{\alpha} \chi^T H_{\alpha} \phi) 
 = i \chi^T [\sum_{\alpha} m_{\alpha} H_{\alpha} , 
   \sum_A \eps^A T_A ] \phi \;  
  \label{variation} 
\eeq
vanishes only when $T_A \sim H_{\alpha}$.   

We eliminate the auxiliary fields $V$ and $\sig$ to 
 obtain the nonlinear Lagrangian.
The equations of motion for $V$ and $\sig$ read
\beq
 && {\del {\cal L} / \del V} 
 = e^V |\phi|^2 - e^{-V} |\chi|^2 - c = 0 \; , 
   \label{EOM-V}\\
 && {\del {\cal L} / \del \sig} 
 = \chi^T \cdot \phi - b= 0 \;, 
   \label{EOM-sig}
\eeq
respectively. 
Setting $X=e^V$ in the first equation, 
 we obtain the algebraic equation 
 $|\phi|^2 X^2 - c X - |\chi|^2 =0$, 
 which can be solved to give 
 $X = (c \pm \sqrt {c^2 + 4|\phi|^2 |\chi|^2})/ (2|\phi|^2)$. 
We thus obtain the K\"ahler potential of the form
\beq
 K = \sqrt{c^2 + 4 |\phi|^2 |\chi|^2 } 
   - c \log \left(c + \sqrt{c^2 + 4 |\phi|^2 |\chi|^2 } \right)
   + c \log |\phi|^2 \;, \label{kahler}
\eeq
where we have chosen the plus sign of the solution 
 for the positivity of the metric. 
This \kahler potential (\ref{kahler}) is still invariant 
 under the $U(1)^{\bf C}$ gauge transformation (\ref{gauge-sym})
 up to a \kahler transformation. 
Fixing a gauge and substituting a solution of (\ref{EOM-sig}), 
 we obtain the Lagrangian of the massive HK sigma model on 
 $T^*{\bf C}P^{n}$ in terms of independent ${\cal N}=1$ superfields.

%By applying an $SU(2)_R$ transformation, we can set 
%i) $c=0$ or ii) $b=0$ 
%without loss of generality. 
We can consider the two cases of i) $c=0$ and ii) $b=0$, 
which are related by an $SU(2)_R$ transformation.

i) $c=0$. Using the $U(1)^{\bf C}$ gauge degree of freedom, 
we can set~\cite{AF2,RT}   
\beq
 \phi = {1 \over {\sqrt{1+ v^T \cdot w}}} \pmatrix{1 \cr v} \;, \hs{5}
 \chi = {b \over {\sqrt{1+ v^T \cdot w}}} \pmatrix{1 \cr w} \;,
   \label{paramet}
\eeq
 where $v$ and $w$ are vectors of $n$ chiral superfields. 
We thus obtain the Lagrangian of the massive HK sigma model on 
 $T^*{\bf C}P^{n}$ whose \kahler potential and 
 superpotential are given by
\beq
&& K = 2 |b| \sqrt{(1+|v|^2)(1+|w|^2) \over |1+ v^T \cdot w|^2} \;,\non
&& W = {b \over 1 + v^T \cdot w} 
      \sum_{{\alpha}=1}^{n} m_{\alpha} (1, w^T) H_{\alpha} 
      \pmatrix{1 \cr v} \;, 
  \label{massive-CPN1}
\eeq
respectively. 

ii) $b=0$.  
In this case, we can set 
\beq
 \phi = \pmatrix{1 \cr v} \;, \hs{5}
 \chi = \pmatrix{- v^T \cdot w \cr w}  \;,
\eeq
by using the $U(1)^{\bf C}$ gauge symmetry.  
We obtain 
\beq
 && K = \sqrt{c^2 + 4 (1 + |v|^2)
           (|v^T \cdot w |^2 + |w|^2 )} \non
 && \hs{10} 
   - c \log \left(c + \sqrt{c^2 
     + 4 (1 + |v|^2)(|v^T \cdot w |^2 + |w|^2)} \right)
   + c \log (1+|v|^2) \;, \non
 && W = \sum_{{\alpha}=1}^{n} m_{\alpha} (- v^T \cdot w, w^T) H_{\alpha} 
       \pmatrix{1 \cr v} \;. \label{massive-CPN2}
\eeq

Since $SU(2)_R$ rotates three complex structures among themselves, 
these two cases ($c=0$ and $b=0$) cannot be related to each other 
by a holomorphic 
%transformation. 
field redefinition.

\if0 %%%%%%%%%%%%%%%%%
\begin{itemize}
\item
Setting $w=0$ in $K$ we obtain $K|_{w=0} = c \log(1+|y|^2)$ 
 which is the \kahler potential of the Fubini-Study metric 
 on ${\bf C}P^{n}$. 
Hence $y$ are the coordinates of ${\bf C}P^{n}$ as 
 the base manifold and $w$ are those of fiber, with 
 the total space being the cotangent bundle.  

\item
The isometry of the manifold is $SU(n)$. 
Its Killing vectors can be calculated as follows: 
The transformation of $SU(n)$ on the linear fields 
 (homogeneous coordinates) $\phi$ and $\chi$, 
 given by (\ref{global}) with $g= e^{i \theta^A T_A}$, 
 changes the gauge fixing condition (\ref{gauge2}) in general. 
Hence, to obtain the transformation law of 
 the inhomogeneous coordinates $y$ and $w$, 
 we need a gauge transformation compensating such a variation. 
It is given by (\ref{gauge-sym}) 
 with a gauge parameter 
 $\Lambda = - \theta^A (T_A)_{11} - \theta^A (T_A)_{1,1+i} y^i$.  
We thus obtain the Killing vectors of the $SU(n)$ isometry given by 
\beq
 k_A 
 = \pmatrix {\delta_A y^i \cr \delta_A w^i} 
 = \pmatrix {
   i (T_A)_{1+i,j} y^j - [i(T_A)_{11} + i(T_A)_{1,1+j} y^j] y^i \cr
   -i (T_A)_{j,1+i} w^j + [i(T_A)_{11} + i(T_A)_{1,1+j} y^j] w^i  
   } \;, \label{Killing}
\eeq
where $\delta_A$ is defined by 
$\delta = \theta^A \delta_A 
= \delta_{\rm global} + \delta_{\rm gauge}$. 

However note that only the transformation by the Cartan generators 
 preserves the superpotential and can be considered as 
 the symmetry of the whole Lagrangian as seen in (\ref{variation}). 

\end{itemize}

\fi %%%%%%%%%%%

%%%%%%%%
\subsection{Massive HK sigma model on Eguchi-Hanson space}
The $n=1$ case of $T^*{\bf C}P^n$,  $\; T^*{\bf C}P^1$ is 
 the Eguchi-Hanson space of the gravitational instanton~\cite{EH}.  
The parameters $b$ and $c$ correspond to 
 the blow up and the deformation of 
 the orbifold singularity in ${\bf C}^2/{\bf Z}_2$, respectively.
In this subsection, 
 we present the explicit form of 
 the Lagrangian in the $c=0$ case.  

Setting $n=1$, 
$ m_1 \equiv  \mu $ 
%$ m_1 \equiv 2 \mu $ 
and
 $H_1 = \1{2} \sig_3$
% $H_1 = T_3 = \1{2} \sig_3$, 
 the superpotential (\ref{massive-CPN1}) becomes 
\beq
 W =   b \mu {1 \over 1 + v w} \;, 
% W = -  b \mu {v w \over 1 + v w} \;, 
\eeq
where we have omitted a constant shift, 
since it disappears under the superspace integral. 

We denote the coordinates of the sigma model manifold by the superfields 
$\ph^i = (v,w)$. 
The \kahler metric $g_{ij^*} = \del_i \del_{j^*} K$, 
where $\del_i$ denotes a differential with respect to $\ph^i$, 
is found to be 
\beq
 g_{ij^*} 
 = {K \over 4} 
 \pmatrix{ 
  \left(1 + {K^2 \over 4|b|^2}\right) \1{(1+|v|^2)^2} & 
  - \1{|1 + vw|^2} {(w - v^*)^2 \over (1+|v|^2) (1+|w|^2)} \cr 
  - \1{|1 + vw|^2} {(v - w^*)^2 \over (1+|v|^2) (1+|w|^2)} & 
  \left(1 + {K^2 \over 4|b|^2}\right) \1{(1+|w|^2)^2}
         } \;,
\eeq
or
\beq
 ds^2 &=& {K \over 4} \left(1 + {K^2 \over 4|b|^2}\right) 
 \left[ {dv dv^* \over (1+|v|^2)^2} 
      + {dw dw^* \over (1+|w|^2)^2}\right] \non
&& - {K \over 4 |1+vw|^2} 
   {(v - w^*)^2 dv dw^* + (w - v^*)^2 dw dv^* 
          \over (1+|v|^2)(1+|w|^2)} \;. \label{metric} 
\eeq
Since the determinant is $\det g_{ij^*} = |b|^2/|1+vw|^4$, 
the inverse of the metric is obtained as 
\beq
 g^{ij^*} = {K |1+ vw|^4 \over 4|b|^2} 
 \pmatrix{ 
  \left(1 + {K^2 \over 4|b|^2}\right) \1{(1+|w|^2)^2} & 
  \1{|1 + vw|^2} {(v - w^*)^2 \over (1+|v|^2) (1+|w|^2)} \cr 
  \1{|1 + vw|^2} {(w - v^*)^2 \over (1+|v|^2) (1+|w|^2)} & 
  \left(1 + {K^2 \over 4|b|^2}\right) \1{(1+|v|^2)^2}
         } \;.  
\eeq
Therefore the scalar potential can be calculated as
\beq
 V &=& g^{ij^*} \del_i W \del_{j^*} W^* \non
 &=& {|\mu|^2 \over 4} K \left[ 
      {|w|^2 \over (1+|w|^2)^2} 
   +  {|v|^2 \over (1+|v|^2)^2} 
   +  {|v|^2 + |w|^2 \over (1+|v|^2)(1+|w|^2)} 
   \right] \; ,\label{potential} 
\eeq
where we have used the same letters with superfields
for their lowest components.
The vacua are given by 
$|v| = |w|=0$ or $|v| = |w| = \infty$. 
  
Next, we show that this scalar potential can be 
rewritten by the norm of the Killing vector 
whose action preserves the superpotential, 
corresponding to the $SU(2)$ generator $\1{2} \sig_3$. 
We note that 
the $SU(2)$ action (\ref{global}) on $\phi$ and $\chi$ breaks 
%does not preserve 
the gauge fixing condition 
of $\chi_1/\phi_1 = b$ in Eq.(\ref{paramet}).   
Hence a compensating $U(1)$ gauge transformation is needed for the 
 $SU(2)$ action on $v$ and $w$ to preserve the gauge fixing condition. 
In the case of $g = e^{i \eps \1{2}\sig_3}$, 
the variation 
$\delta_{\eps} (\chi_1/\phi_1) = i \eps (\chi_1/\phi_1)$ 
 should be compensated by (\ref{gauge-sym}) with 
 $\Lambda= -\eps/2$. 
Therefore we find $\delta_3 v \equiv \delta_{\eps} v 
+ \delta_{\Lambda} v = - i \eps v $ 
and $\delta_3 w \equiv \delta_{\eps} w 
+ \delta_{\Lambda} w = i \eps w$.  
We thus obtain the Killing vector for $\1{2}\sig_3$, 
given by 
\beq
 k_3^i = \frac{1}{\eps}\pmatrix{\delta_3 v \cr \delta_3 w}
 = \pmatrix{- i v \cr  i w} \label{eq:killing}\;.
\eeq
Using this Killing vector, we find  
\beq
 V =  |\mu|^2 g_{ij^*} k_3^i k_3^{*j} \;. 
% V = \1{4} |\mu|^2 g_{ij^*} k_3^i k_3^{*j} \;. 
\eeq

We can consider the projection map from the bundle $T^*{\bf C}P^1$ 
 to the base manifold ${\bf C}P^1$.
It is given by $v = w^*$~\cite{AF2}. 
By this map, the metric (\ref{metric}) 
 is mapped into the Fubini-Study metric 
 on ${\bf C}P^1$ 
%%%%%%%%%%%%%%%%%%%%%%
\beq
 ds^2|_{v=w^*} = {2|b| dv dv^* \over (1+|v|^2)^2} \;, \label{CP1}
\eeq
%%%%%%%%%%%%%%%%%%%%%%
and the potential (\ref{potential}) is reduced to 
\beq
 V|_{v=w^*} = {2 |b| |\mu|^2 |v|^2 \over (1+|v|^2)^2} \label{CP1V}\;,
\eeq
which coincides with the one of 
the massive ${\bf C}P^1$ model~\cite{AT}.

%%%%%%%%%%%%%%%%%%%%%%%%%%%%%%%%%%%%%%%%%%%
\subsection{BPS equation and its solution}
In this subsection, 
we construct the BPS domain wall 
in the massive $T^*{\bf C}P^1$ model.
The SUSY transformation on the fermion is given by 
\beq
 \delta_{\epsilon} \psi^i 
 = i \sqrt 2 \sigma^\mu {\bar{\epsilon}} \partial_\mu \ph^i 
 + \sqrt 2 \epsilon F^i \;. 
\eeq
Let us choose $y=x^2$ as the spatial direction perpendicular to
the BPS domain wall. 
Without loss of generality, we can require the direction of 
preserved SUSY as 
\beq
 e^{i\alpha} \sig^2 \bar\epsilon =i \epsilon 
\label{N1harfSUSY}
\eeq
with a phase factor $e^{i\alpha}$ to be determined later. 
Then the BPS equations are given by \cite{OINS} 
\beq
 \del_2 \ph^i = - e^{i\alpha} 
g^{ij^*} \del_{j^*} W^* \;,  
\eeq
where the both sides are evaluated at classical fields. 
In the case of the massive $T^*{\bf C}P^1$ model, 
 after eliminating the auxiliary fields, these BPS equations 
reduce to 
\beq
 && \del_2 v = e^{i\alpha}  {\mu^* \over 4 b} K (1 + vw)^2 
  \left[{|1+vw|^2 + (1 +|v|^2)(1+|w|^2) 
        \over |1+vw|^2 (1+|w|^2)^2} w^*  
  + { (v-w^*)^2 v^* \over |1+vw|^2 (1+|v|^2)(1+|w|^2) } \right], \non
 && \del_2 w =  e^{i\alpha}  {\mu^* \over 4 b} K (1 + vw)^2 
  \left[{|1+vw|^2 + (1 +|v|^2)(1+|w|^2) 
        \over |1+vw|^2 (1+|v|^2)^2} v^*  
      + {(w-v^*)^2 w^* \over |1+vw|^2 (1+|v|^2)(1+|w|^2) } \right].
 \non
\label{eq:BPS_vw}
\eeq 
%%%%%%%%%%%%%%%%%%%%%%%%%%%%%%%%%%%%%%%
Now we must choose the phase $e^{i\alpha}$ to absorb 
the phase of the parameter
\footnote{
 For simplicity, we choose $\mu$ to be real positive 
 in the following.
} $\mu^*/b$ 
\begin{equation}
e^{i\alpha}{\mu^* \over b}=\left|{\mu \over b}\right|. 
\end{equation}
By subtracting the complex conjugate of the second equation from 
the first one in Eq.(\ref{eq:BPS_vw}), we obtain 
\begin{eqnarray}\label{eq:BPS_v-w*}
%\begin{equation}
 &\!\!\! &\!\!\!
{\partial (v-w^*) \over \partial y}=
\left|{\mu \over b}\right|{K \over 4}
\left[
\left\{\left({1+vw \over |1+vw|}\right)^2v^*
-\left({1+v^*w^* \over |1+vw|}\right)^2w\right\}
 {(v-w^*)^2 \over (1+|v|^2)(1+|w|^2)}
\right.
 \\
 &\!\!\! 
%\times
+
 &\!\!\!
\left.
\left\{\left({1+vw \over |1+vw|}\right)^2{w^* \over (1+|w|^2)^2}
-\left({1+v^*w^* \over |1+vw|}\right)^2{v \over (1+|v|^2)^2}\right\}
\left\{|1+vw|^2+(1+|v|^2)(1+|w|^2)\right\}
\right], 
\nonumber
\end{eqnarray}
whose right-hand side vanishes for $v=w^*$. 
The BPS equation (\ref{eq:BPS_v-w*}) 
dictates that $v=w^*$ is valid for arbitrary $y$, 
if an initial condition $v=w^*$ is chosen at some $y$. 
Since we can choose the initial condition $v=w^*$ at $y=-\infty$, 
we find the BPS equations (\ref{eq:BPS_vw}) simply reduce to 
\beq
 \del_2 v =   |\mu| v \;,
\eeq
which is the BPS equation on the submanifold ${\bf C}P^1$ (\ref{CP1})
 defined by $v = w^*$. 
Therefore we obtain a BPS wall configuration connecting two vacua 
$v=w^*=0$ at $y=-\infty$ to $v=w^*=\infty$ at $y=\infty$ 
along $v=w^*$ with a constant phase $e^{i\varphi_0}$ 
\beq
 v = w^* = e^{|\mu| (y+y_0)}e^{i\varphi_0}\;,    
 \label{solution} 
\eeq
where $y_0$ is also a constant representing the position of the wall. 
Thus we find two collective coordinates (zero modes) corresponding 
to the spontaneously broken translation ($y_0$) and $U(1)$ symmetry 
($\varphi_0$). 

We can show that BPS solution (\ref{solution}) coincides with 
 that derived in component formalism \cite{GTT1}
 through the following field redefinition
 \footnote{Actually, using this field redefinition, one can show 
 that massive ${\bf C}P^1$ model corresponds to the
 truncated model of massive $T^*{\bf C}P^1$ model in  
 component formalism given in Ref.~\cite{GTT1}. 
 We discuss the truncated 
 model further in section 4.}
 $v\rightarrow X, \varphi$
%%%%%%%%%%%%%%%%%%%%%%%%%%%%%%%%%%%%
\begin{eqnarray}
 %&
 v \equiv e^{u + i \varphi}, 
\qquad 
%\label{vbase} & \\ & \rightarrow
  X=|b|\tanh u,
%~~\varphi=\varphi,
%& 
\end{eqnarray}
%%%%%%%%%%%%%%%%%%%%%%%%%%%%%%%%%%%%
where $u,~\varphi$ and $X$ are real scalar fields.
%The wall solution is recognised $u(y)=\mu(y+y_0)$ and $\varphi(y)=0$. 
After the field redefinition, the theory of massive ${\bf C}P^1$
 model is described by $X$ and $\varphi$, and 
 the wall solution (\ref{solution}) is mapped to
%%%%%%%%%%%%%%%%%%%%%%%%%%%%%%%%%%%%
\begin{eqnarray}
 X=|b|\tanh |\mu|(y + y_0), \qquad 
\varphi = \varphi_0. \label{GPT-sol}
\end{eqnarray}
%%%%%%%%%%%%%%%%%%%%%%%%%%%%%%%%%%%%
This solution coincides with that derived in 
 Ref.~\cite{GTT1}.

%Finally we would like to mention about the half SUSY condition for second
% SUSY......
%%%%%%%%%%%%%%%%%%%%%%%%%%%%%%%%%%%%%%%%%%%%%%%%%%%%%%%%%%%%%%%
%%%%%%%%%%%%%%%%%%%%%%
%                    %
% Section 3 HSF      %           
%                    %
%%%%%%%%%%%%%%%%%%%%%%
%\include{hss_section1}
%%%%%%%%%%%%%%%%%%%%%%%%%%%%%%%%%%%%%%%%%%%%%%%%%
% HSF
%%%%%%%%%%%%%%%%%%%%%%%%%%%%%%%%%%%%%%%%%%%%%%%%%
%
% Last Modified 8/17/2002 M.Arai hss_section.tex
%
% Last Modified 8/24/2002 M.Arai hss_section1.tex 
%
%%%%%%%%%%%%%%%%%%%%%%%%%%%%%%%%%%%%%%%%%%%%%%%%%

\def\aarr#1{\addtocounter{mycount#1}{1}\begin{eqnarray}}
\newcommand{\ptheta}{\theta^+}
\newcommand{\bptheta}{{\bar{\theta}}^+}
\renewcommand{\baselinestretch}{1.4}
\textwidth 150mm
\textheight 220mm

\makeatletter
\renewcommand{\theequation}{%
   \thesection.\arabic{equation}}
 \@addtoreset{equation}{section}
\makeatother

%\begin{document}
\section{HSF 
%Harmonic Superspace Formalism 
and domain wall solution}\label{sc:HSF}
In this section,  we describe the massive HK sigma model on
 $T^*{\bf C}P^{n}$ in the HSF,
 and examine BPS equations of the $n=1$ case,
 massive $T^*{\bf C}P^{1}$ model.

As we discussed in the previous section, 
 domain wall solutions can be obtained in the 
 $T^*{\bf C}P^{1}$ case as (\ref{solution}).
As in the ${\cal N}=1$ SUSY theory, we can obtain the BPS equations from
 the SUSY transformations for fermions imposing half SUSY condition,
 and by eliminating auxiliary fields.
The main difference between ${\cal N}=1$ formalism 
and ${\cal N}=2$ formalism (HSF) is that
 there is an infinite set of auxiliary fields in the harmonic superfield,
 while there is single auxiliary field $F^i$ for each  
 chiral superfield in the ${\cal N}=1$ superfield formalism.
As a result, BPS conditions contain an infinite set of auxiliary fields
 in addition to physical fields. 
To obtain the BPS equations, the infinite set of the auxiliary
 fields should be eliminated by using the solution of 
 the equations of motion for auxiliary fields.
We call these solutions ``on-shell condition''.
After substituting the on-shell condition and the half SUSY 
 condition into the SUSY transformation of fermions, the BPS equations 
 can be obtained.

In the following, we first describe the action of the
 massive $T^*{\bf C}P^{n}$ model in the HSF, and 
 briefly describe
 %review 
 the $n=1$ case.
Next we derive the equations of motion for auxiliary fields and
 show how to eliminate the infinite set of auxiliary fields.
Then, it is shown that the BPS equations are obtained by 
 using the on-shell condition.
Finally we solve the BPS equations and show that the solution
 coincides with Eq.~(\ref{GPT-sol}).
Notation we use in this section is summarized in Appendix \ref{sc:nothss}.
%%%%%%%%%%%%%%%%%%%%%%%%%%%%
%
% T^* CP^{n-1}
%
%%%%%%%%%%%%%%%%%%%%%%%%%%%%
\subsection{Massive HK sigma model on $T^*{\bf C}P^{n}$} \label{sc:MTSCP1HSS}

We can easily describe the massive HK sigma model on 
 $T^*{\bf C}P^{n}$ 
 in the HSF by considering the action in terms of the 
 ${\cal N}=1$ superfield formalism (\ref{linear}). 
We consider %The action is described by 
($n+1$)-hypermultiplets 
 $\phi_a^+~(a=1,\dots,n+1)$ in the fundamental representation
 of $SU(n+1)$ 
which transform under 
$U(1)$ gauge transformation
%$U(1)$ gauge transformations 
with unit charge 
%%%%%%%%%%%%%%%%%%%%%%%%%%%%%
\begin{eqnarray}
\phi_a^+ \rightarrow e^{-i \lambda(\zeta_A,u)} \phi_a^+,
\end{eqnarray}
%%%%%%%%%%%%%%%%%%%%%%%%%%%%%
where $\lambda$ is the real analytic superfield 
representing the gauge transformation parameter, 
%with vanishing $U(1)$ charge,
 and also 
%by 
the vector multiplet transforming under the $U(1)$ gauge
 transformation as 
%%%%%%%%%%%%%%%%%%%%%%%%%%%%%
\begin{eqnarray}
  \delta V^{++} &=& D^{++} \lambda(\zeta_A,u). \label{gthss3}
\end{eqnarray}
%%%%%%%%%%%%%%%%%%%%%%%%%%%% 
Then, the action of the massive $T^*{\bf C}P^{n}$ model is
 described as 
%%%%%%%%%%%%%%%%%%%%%%%%%%%%%
\begin{eqnarray}
 S = -\displaystyle\int d\zeta_A^{(-4)} du \sum_{a=1}^{n+1}  
     \left\{\widetilde{\phi_a^{+}}(D^{++}+i V^{++})\phi_a^{+}
           +\xi^{++} V^{++}  
           \right\}, \label{TSCPN}
\end{eqnarray}
%%%%%%%%%%%%%%%%%%%%%%%%%%%%%
where 
 $\xi^{++}=\xi^{(ij)}u_{(i}^+u_{j)}^+$ 
 is the coefficient of the 
 FI term which is the $SU(2)_R$ triplet.
Harmonic variables are denoted as 
$u_i^{\pm}$, $i=1,2$ being $SU(2)_R$ indices. 
See Appendix \ref{sc:nothss} for details. 
The integral measure $d\zeta_A^{(-4)}$ is the analytic measure 
 which is defined by 
 $d\zeta_A^{(-4)}=d^4 x_A d^2 \theta^{+} d^2 \bar{\theta}^{+}$.
The covariant derivative $D^{++}$ is defined as
%%%%%%%%%%%%%%%%%%%%%%%%%%%%%
\begin{eqnarray}
 D^{++}=\partial^{++}-2 i \theta^+ \sigma^\mu {\bar{\theta}}^+ 
        \partial_\mu^A - (\theta^{+2}\bar{Z}-{\bar{\theta}}^{+2}Z), 
        \label{covdhss}
\end{eqnarray}
%%%%%%%%%%%%%%%%%%%%%%%%%%%%%
where $\partial^{++}$ is harmonic differential defined by 
 $\partial^{++}=u_i^+\frac{\partial}{\partial u_i^-}$,
 and $\partial_\mu^A$ is the spacetime derivative in analytic basis.
The central charge is denoted as $Z$ 
whose eigenvalue is given as
 \footnote{Since the central charge is defined by 
 $Z=-i(\partial_5+i \partial_6)$, the solution 
 of Eq.~(\ref{eq:hsseigen}) depends on extra spacetime
 $x_5$ and $x_6$. But the action does not depend on 
 them. See Ref.~\cite{Ivanov2} in detail.}
%%%%%%%%%%%%%%%%%%%%%%%%%%%%%
\begin{eqnarray}
 Z \phi_a^+ = \sum_{\alpha=1}^{n} \sum_{b=1}^{n+1} 
 m_\alpha ({H_\alpha})_{ab} \phi_b^{+}, \label{eq:hsseigen} 
\end{eqnarray}
%%%%%%%%%%%%%%%%%%%%%%%%%%%%%
where $m_\alpha$ is complex mass parameters, and
 $H_\alpha$ are the diagonal generators of the Cartan subalgebra of 
 $SU(n+1)$ as in (\ref{linear}).
The central charge $Z$ vanishes 
 for fields neutral under $SU(n+1)$ such as $V^{++}$. 
In the limit of $m_\alpha=0$,
 the action (\ref{TSCPN}) becomes massless $T^*{\bf C}P^{n}$
 model whose isometry is $SU(n+1)$. 
The mass term explicitly 
 breaks $SU(n+1)$ into $U(1)^{n}$.
These features are identical to 
the case of ${\cal N}=1$ formalism. 
%%%%%%%%%%%%%%%%%%%%%%%%%%%%%%%%%%%%%
%
% T^*CP^1
%
%%%%%%%%%%%%%%%%%%%%%%%%%%%%%%%%%%%%%
\subsection{Massive HK sigma model on Eguchi-Hanson space}
In the following, we focus on the $n=1$ case, massive 
 HK sigma model on $T^*{\bf C}P^1$.
Here we follow the original notation introduced in 
 Ref.~\cite{ivanov-EH}, which uses $O(2)$ gauge invariant form 
instead of the $U(1)$ in Eq.(\ref{TSCPN}). 
It is described by 
  \footnote{The action (\ref{TSCPN}) with $n=1$ and (\ref{eq:EHhss}) 
    is related by
    $\phi_1^+=\frac{1}{\sqrt{2}}(q_1^+ - i q_2^+),
    ~\phi_2^{+\prime}=\frac{1}{\sqrt{2}}(q_1^+ + i q_2^+)$
    with the identification 
     ${\widetilde{\phi_2^{+}}}\equiv \phi_2^{+\prime}~
     (\phi_2^+=-{\widetilde{\phi_2^{+\prime}}})$.
  }
%%%%%%%%%%%%%%%%%%%%%%%%%%%%%
\begin{eqnarray}
 S=-\displaystyle\int d\zeta_A^{(-4)}du \left({\widetilde{q_1^+}} D^{++}q_1^+
   +{\widetilde{q_2^+}}D^{++}q_2^+ +V^{++}({\widetilde{q_1^+}} q_2^+ 
   -{\widetilde{q_2^+}}q_1^+ +\xi^{++})\right), \label{eq:EHhss}
\end{eqnarray}
%%%%%%%%%%%%%%%%%%%%%%%%%%%%%
where the central charge $Z$ satisfies the following eigenvalue 
 equation which is obtained by using field redefinition, and
 taking $n=1, m_1=\mu \in {\rm \bf R}$ and 
 $H_1 %= T_3 
= \1{2} \sig_3$ 
% $H_1=\frac{1}{2}T_3$ 
in (\ref{eq:hsseigen})
%%%%%%%%%%%%%%%%%%%%%%%%%%%%%
\begin{eqnarray}
 Z q_a^+ = \frac{\mu}{2} q_a^+,
\end{eqnarray}
%%%%%%%%%%%%%%%%%%%%%%%%%%%%%
where we take the complex mass parameter $\mu$ to be real for simplicity 
% (in detail, see
% Appendix \ref{ap:BPSeqHSS})
.
The action (\ref{eq:EHhss}) is invariant under $O(2)$ gauge
 transformation 
%%%%%%%%%%%%%%%%%%%%%%%%%%%%%
\begin{eqnarray}
 \delta q_1^+ &=& -\lambda(\zeta_A,u) q_2^+, \label{gthss1} \\
 \delta q_2^+ &=& \lambda(\zeta_A,u) q_1^+, \label{gthss2}
\end{eqnarray}
%%%%%%%%%%%%%%%%%%%%%%%%%%%%%
 and (\ref{gthss3}).

To write down the component action 
 (\ref{eq:EHhss}), we derive the equations of motion
  \footnote{In this section, we express the action using harmonic 
   superfields with constraints instead of independent ones, 
   in contrast to 
   the ${\cal N}=1$ case (\ref{massive-CPN1}) and (\ref{massive-CPN2}) 
   which were obtained after eliminating the Lagrange multipliers. 
   Action can be expressed by independent harmonic superfields as 
   in Refs.~\cite{Ivanov2,valent}.
   However, we will solve the constraint 
   and gauge away redundant degrees of freedom, 
   after writing down the on-shell action.
  }.
Varying (\ref{eq:EHhss}) with respect to the superfields $q_a^+$, 
 (and their conjugate) and $V^{++}$ yield the equations of motion,
%%%%%%%%%%%%%%%%%%%%%%%%%%%%%
\begin{eqnarray}
D^{++} q_1^+ + V^{++} q_2^+ &=& 0, \label{eq:emhss1} \\
D^{++} q_2^+ - V^{++} q_1^+ &=& 0, \label{eq:emhss2} \\ 
{\widetilde{q_1^+}} q_2^+ - {\widetilde{q_2^+}} q_1^+ +\xi^{++} &=& 0, 
 \label{eq:emhss3}
\end{eqnarray}
%%%%%%%%%%%%%%%%%%%%%%%%%%%%%
where (\ref{eq:emhss1}) and (\ref{eq:emhss2}) include kinematical
 and dynamical parts, and (\ref{eq:emhss3}) is a constraint.
The auxiliary fields are eliminated by using the solutions of 
 the kinematical part of
 Eqs.~(\ref{eq:emhss1}) and (\ref{eq:emhss2}).
To derive the kinematical part of 
 equations of motion, we substitute the component
 expansion (\ref{eq:exp1}) and (\ref{eq:exp2}) for
 the analytic superfields $q_a^+$ and $V^{++}$
 into (\ref{eq:emhss1}) and (\ref{eq:emhss2})
 \footnote{Here we take the Wess-Zumino gauge 
where the gauge transformation is not complexified 
but real~(see (\ref{eq:exp2})).}.
Then, one obtains the equations of motion as
 (\ref{appeq11})-(\ref{appeq27}) for 
 the Grassmann coefficients in (\ref{eq:exp1}) and (\ref{eq:exp2}),
 and one can solve easily the kinematical part 
 (\ref{appeq11})-(\ref{appeq14}) and (\ref{appeq21})-(\ref{appeq24}).
The solutions are given by 
%%%%%%%%%%%%%%%%%%%%%%%%%%%%%
\begin{eqnarray}
 F_a^+(x_A,u) & = & f_a^i(x_A) u_i^+, \label{solhss1} \\
  \psi_a(x_A,u) & = & \psi_a(x_A),
 ~~{\bar{\varphi}}_a(x_A,u)={\bar{\varphi}}_a(x_A),
  \label{solhss2} \\
 A_{1\mu}^-(x_A,u) &=& 2 (\partial_\mu^A f_1^i + V_\mu f_2^i)(x_A) u_i^-, 
  \label{solhss3} \\ 
 A_{2\mu}^-(x_A,u) &=& 2 (\partial_\mu^A f_2^i - V_\mu f_1^i)(x_A) u_i^-, 
  \label{solhss4} \\
 M_1^-(x_A,u) &=& -\left(\bar{M}_v f_2^i - \frac{\mu}{2} 
f_1^i\right)(x_A) u_i^-, 
  \label{solhss5} \\
 M_2^-(x_A,u) &=&  \left(\bar{M}_v f_1^i + \frac{\mu}{2} 
f_2^i \right)(x_A) u_i^-, 
  \label{solhss6} \\
 N_1^-(x_A,u) &=&  -\left(M_v f_2^i + \frac{\mu}{2} f_1^i \right)(x_A) u_i^-, 
                  \label{solhss7} \\
 N_2^-(x_A,u) &=&  \left(M_v f_1^i - \frac{\mu}{2} f_2^i\right)(x_A) u_i^-. 
                  \label{solhss8}  
\end{eqnarray}
%%%%%%%%%%%%%%%%%%%%%%%%%%%%%
Note that the infinite set of auxiliary fields in the harmonic
 expansion are eliminated and the physical fields $f_a^i$, 
 $\psi_a$, ${\bar{\varphi}}_a$ and
 the Lagrange multipliers $M_v$, $V_\mu$ are left.
The latter are eliminated by using algebraic equations as 
 will be mentioned later.

At this stage, we can write down the component action.
In the following, we focus on the bosonic part of the action
 in order to obtain the equations of motion for Lagrange
 multipliers which are necessary to derive the BPS equations
 \footnote{We write down the full on-shell action including fermions
 in Appendix \ref{sc:TSCP1HSS}.}. 
Substituting (\ref{eq:exp1}),~(\ref{eq:exp2}) and 
 (\ref{solhss1})-(\ref{solhss8})
 into the action (\ref{eq:EHhss}), and integrating Grassmann variables
 and the harmonic variable, the bosonic part of the action 
 becomes 
%%%%%%%%%%%%%%%%%%%%%%%%%%%%%
\begin{eqnarray}
 S_{\rm boson} &=& -\displaystyle\int d^4 x_A 
  \Bigg{\{}(\partial_A^\mu f_1^i + V^\mu f_2^i)
  (\partial_\mu^A {\bar{f}}_{1i} + V_\mu {\bar{f}}_{2i}) 
  \nonumber \\
 & & +
  (\partial_A^\mu f_2^i - V^\mu f_1^i)
  (\partial_\mu^A {\bar{f}}_{2i} - V_\mu {\bar{f}}_{1i}) 
  \nonumber \\
 & & -\frac{1}{2}\left({\bar{M}}_v {\bar{f}}_1^i- \frac{\mu}{2} {\bar{f}}_2^i\right)
                 \left(M_v f_{1i} - \frac{\mu}{2} f_{2i}\right)
     -\frac{1}{2}\left({\bar{M}}_v {\bar{f}}_2^i + \frac{\mu}{2} {\bar{f}}_1^i\right)
                 \left(M_v f_{2i} + \frac{\mu}{2} f_{1i}\right)
  \nonumber \\
 & & -\frac{1}{2}\left(M_v {\bar{f}}_1^i + \frac{\mu}{2} {\bar{f}}_2^i\right)
                 \left({\bar{M}}_v f_{1i} + \frac{\mu}{2} f_{2i}\right)
     -\frac{1}{2}\left(M_v {\bar{f}}_2^i - \frac{\mu}{2} {\bar{f}}_1^i\right)
                 \left({\bar{M}}_v f_{2i} - \frac{\mu}{2} f_{1i}\right) \nonumber \\
 & & +\frac{1}{3}D_{v(ij)}(-{\bar{f}}_1^{(i}f_2^{j)}
     +{\bar{f}}_2^{(i}f_1^{j)}+\xi^{(ij)}) 
     \Bigg{\}}
     \label{eq:bahss}.
\end{eqnarray}
%%%%%%%%%%%%%%%%%%%%%%%%%%%%
%There are still auxiliary fields $M_v$ and $V^\mu$ in the 
% vector multiplet.
Equations of motion for the auxiliary fields $M_v$ and $V^\mu$ are given by
%%%%%%%%%%%%%%%%%%%%%%%%%%%%
\begin{eqnarray}
 M_v &=& -{\bar{M}}_v = -\frac{\mu}{2}\frac{(f_1^i {\bar{f}}_{2i} 
                               -f_2^i{\bar{f}}_{1i})}
          {f_1^i {\bar{f}}_{1i} + f_2^i {\bar{f}}_{2i}}, 
         \label{eq:vecscalar1}\\
 2 V^\mu &=& \frac{-(\partial^\mu_A {\bar{f}}_{1i} f_2^i
                   -{\bar{f}}_{1i}\partial^\mu_A f_2^i
                   -\partial^\mu_A {\bar{f}}_{2i} f_1^i
                   +{\bar{f}}_{2i} \partial_A^\mu f_1^i)}
                 {f_1^i {\bar{f}}_{1i} + f_2^i {\bar{f}}_{2i}}. 
         \label{eq:vecscalar2}
\end{eqnarray}
%%%%%%%%%%%%%%%%%%%%%%%%%%%%
Substituting (\ref{eq:vecscalar1}) and (\ref{eq:vecscalar2}) into 
 (\ref{eq:bahss}), we finally obtain the bosonic part of the action
%%%%%%%%%%%%%%%%%%%%%%%%%%%%
\begin{eqnarray}
 S_{\rm boson} &=& \displaystyle \int d^4 x_A 
               \Bigg(-\partial_A^\mu f_1^i \partial_\mu^A {\bar{f}}_{1i}
                     -\partial_A^\mu f_2^i \partial_\mu^A {\bar{f}}_{2i}
               \nonumber \\
           & & +\frac{(\partial^\mu_A {\bar{f}}_{1i} f_2^i
                   -{\bar{f}}_{1i}\partial^\mu_A f_2^i
                   -\partial^\mu_A {\bar{f}}_{2i} f_1^i
                   +{\bar{f}}_{2i} \partial_A^\mu f_1^i)^2}
                    {4 (f_1^i {\bar{f}}_{1i} + f_2^i {\bar{f}}_{2i})} 
               \nonumber \\
           & & -\frac{1}{3}D_{(ij)}(-{\bar{f}}_1^{(i}f_2^{j)}
               +{\bar{f}}_2^{(i}f_1^{j)}+\xi^{(ij)}) 
               -V(f_1,f_2)
               \Bigg{)}, \label{bosonachss} \\
 V(f_1,f_2) &=& \frac{\mu^2}{4}
               \frac{1}{f_1^i {\bar{f}}_{1i} + f_2^i {\bar{f}}_{2i}}
               \left\{-
%(
|f_1^i {\bar{f}}_{2i} - f_2^i {\bar{f}}_{1i}
%)
|^2
                +(f_1^i {\bar{f}}_{1i} + f_2^i {\bar{f}}_{2i})^2 \right\}.
               \label{pothss}
\end{eqnarray} 
%%%%%%%%%%%%%%%%%%%%%%%%%%%%
It was proved that the target metric for the four independent 
 bosonic fields is just the Eguchi-Hanson 
 metric \cite{CF,ivanov-EH,valent}.
To see that the dimension of the physical boson manifold 
 equals four, one should take into account that the constraint in
 (\ref{bosonachss}) 
%%%%%%%%%%%%%%%%%%%%%%%%%%%%
\begin{eqnarray}
 -{\bar{f}}_1^{(i}f_2^{j)}+{\bar{f}}_2^{(i}f_1^{j)}+\xi^{(ij)}
 =0 \label{consthss}
\end{eqnarray}
%%%%%%%%%%%%%%%%%%%%%%%%%%%%
 eliminates three out of the original eight bosonic
 degrees of freedom, while one more degree of freedom is gauged away
 by the $O(2)$ gauge invariance.
In the end of this section, we explicitly show that the kinetic term in 
 the action (\ref{bosonachss}) corresponds to the nonlinear sigma 
 model with Eguchi-Hanson metric.

Let us also note that the theory has discrete SUSY vacua 
 \footnote{The scalar potential (\ref{pothss}) was originally derived in 
   Ref.~\cite{Ketov}.
  It was argued that supersymmetry is spontaneously broken
   in contrast to our result of partial SUSY conservation.
  As is shown at the end of this section, there exist two 
   discrete SUSY vacua in the theory. 
  }.
After describing the potential (\ref{pothss}) in terms of the four 
 independent variables, 
 it is found that the potential (\ref{pothss}) corresponds 
 to one which was originally derived 
 in Ref.~\cite{GTT1} (see (\ref{nlsm_lag})), and  
 that there exist two SUSY discrete vacua. 
These SUSY vacua are understood as the fixed points of the Killing
 vector,  as will be seen at the end of this section.
%%%%%%%%%%%%%%%%%%%%%%%%%%%%
%
% BPS equations and solutions
%
%%%%%%%%%%%%%%%%%%%%%%%%%%%%
\subsection{BPS equation and its solutions} \label{BPShss}
In this subsection, we derive the BPS equations, but
 we give here only an outline how to derive the BPS equations.
Detailed derivation is given in Appendix \ref{ap:BPSeqHSS}.

In order to obtain the BPS equations, we have to derive the
 SUSY transformations for fermions.
They can be derived in a model independent way although 
 an infinite set of the auxiliary fields are involved.
In our case, the SUSY transformations can be derived for 
 $\lambda^-(\zeta_A,u)$ in the vector multiplet $V^{++}(\zeta_A,u)$, 
 and $\psi_a(\zeta_A,u),
 ~{\bar{\varphi}}_a(\zeta_A,u),~\xi_a^{--}(\zeta_A,u)$ 
 and ${\bar{\chi}}_a^{--}(\zeta_A,u)$ in the hypermultiplets 
 $q_a^+(\zeta_A,u)$, 
 since the action (\ref{eq:EHhss}) is described by those superfields. 
However, we do not have to derive the SUSY transformations for
 all fermionic components because most fermions are auxiliary
 fields.
Actually, since the bosonic part of the action (\ref{bosonachss})
 is described by the on-shell component $f_a^i(x_A)$,
 it is enough to derive the SUSY transformations for their superpartners
 $\psi_a(x_A)$ and ${\bar{\varphi}}_a(x_A)$ which are the first order 
 components
 in Grassmann expansion of the analytic superfields (\ref{eq:exp1})
 \footnote{Since the scalars $f_a^i(x_A)$ are $SU(2)_R$
  doublets in the Fayet-Sohnius hypermultiplets, their superpartners
  are $SU(2)_R$ singlet, $\psi_a(x_A)$ and ${\bar{\varphi}}_a(x_A)$.
 Actually, it is found that full on-shell action is described by 
  $f_a^i(x_A)$, $\psi_a(x_A)$ and ${\bar{\varphi}}_a(x_A)$ 
  (see Appendix \ref{sc:TSCP1HSS}).
 Alternatively, it is found that the fields $\xi_a^{--}$ and 
  ${\bar{\chi}}_a^{--}$ are auxiliary fields since they do not have 
  the kinetic term (see the equations of motion 
  (\ref{appeq15})-(\ref{appeq17}) and 
  (\ref{appeq25})-(\ref{appeq27})).}.

Recall that the Wess-Zumino gauge is chosen in our case. 
Since SUSY transformations break the Wess-Zumino gauge,
 one has to pull back to the Wess-Zumino gauge by using the $O(2)$
 gauge transformation (\ref{gthss3}).
As a result, SUSY transformation $\hat{\delta}$ in the Wess-Zumino gauge 
 is defined as  $\hat{\delta}=\delta_S+\delta_G$
 where $\delta_S$ and $\delta_G$ are the original SUSY 
 transformation and $O(2)$ gauge transformation, respectively, 
in order to preserve the Wess-Zumino gauge. 
The gauge parameters $\lambda(\zeta_A,u)$ in $O(2)$ gauge 
 transformation are determined
 so as to keep the Wess-Zumino gauge ((\ref{gfhss1})-(\ref{gfhss7})).
Putting altogether, the SUSY transformations for on-shell fermions
 are derived as (\ref{fsv1})-(\ref{fsv4}).
Substituting the on-shell condition (\ref{solhss1})-(\ref{solhss8})
 and the half SUSY condition (\ref{eq:half-SUSY}) into 
 (\ref{fsv1})-(\ref{fsv4}),
 we obtain the BPS equations, 
%%%%%%%%%%%%%%%%%%%%%%%%%%%%
\begin{eqnarray}
 \hat \delta\psi_1 &=& \sqrt{2} \epsilon_1 \left\{({\bar{M}}_v - V_2) f_2^1
               -\left(\frac{\mu}{2} +\partial_2^A\right) f_1^1 \right\} 
\nonumber \\
              & & ~~~~~~~~~~
               +\sqrt{2} \epsilon_2 \left\{({\bar{M}}_v + V_2) f_2^2
               -\left(\frac{\mu}{2} -\partial_2^A\right) f_1^2\right\}=0, 
               \label{eq:BPSeqhss1} \\
 \hat \delta{\bar{\varphi}}_1 
              &=& \sqrt{2} {\bar{\epsilon}}_1 
               \left\{(M_v + V_2) f_2^1
                +\left(\frac{\mu}{2} +\partial_2^A\right) f_1^1 \right\} 
\nonumber \\
              & & ~~~~~~~~~~
               +\sqrt{2} {\bar{\epsilon}}_2 \left\{(M_v - V_2) f_2^2
               +\left(\frac{\mu}{2} -\partial_2^A\right) f_1^2\right\}=0, 
               \label{eq:BPSeqhss2} \\
 \hat \delta\psi_2 &=& \sqrt{2} \epsilon_1 \left\{-({\bar{M}}_v - V_2) f_1^1
                 -\left(\frac{\mu}{2}+\partial_2^A\right) f_2^1 \right\} 
\nonumber \\
              & & ~~~~~~~~~~
               +\sqrt{2} \epsilon_2 \left\{-({\bar{M}}_v + V_2) f_1^2
               -\left(\frac{\mu}{2} -\partial_2^A\right) f_2^2\right\}=0, 
               \label{eq:BPSeqhss3} \\
 \hat \delta{\bar{\varphi}}_2 
              &=& \sqrt{2} {\bar{\epsilon}}_1 
               \left\{-(M_v + V_2) f_1^1
                 +\left(\frac{\mu}{2}+\partial_2^A\right) f_2^1 \right\} 
\nonumber \\
              & & ~~~~~~~~~~
               +\sqrt{2} {\bar{\epsilon}}_2 \left\{-(M_v - V_2) f_1^2
               +\left(\frac{\mu}{2}-\partial_2^A\right) f_2^2\right\}=0. 
               \label{eq:BPSeqhss4}
\end{eqnarray}
%%%%%%%%%%%%%%%%%%%%%%%%%%%%
To satisfy the BPS equations (\ref{eq:BPSeqhss1})-(\ref{eq:BPSeqhss4}), 
 all coefficients of $\epsilon_i$ must vanish, namely there are eight 
 BPS equations.
However, using the relation $M_v=-{\bar{M}}_v$ (see
(\ref{eq:vecscalar1})),
 we find that only four equations are independent;
%%%%%%%%%%%%%%%%%%%%%%%%%%%%
\begin{eqnarray}
 (M_v + V_2) f_2^1 + \left(\frac{\mu}{2} + \partial_2^A\right) f_1^1 &=& 0, 
   \label{eq:BPSeqhss5} \\
 (M_v - V_2) f_2^2 + \left(\frac{\mu}{2} - \partial_2^A\right) f_1^2 &=& 0, 
   \label{eq:BPSeqhss6} \\
 -(M_v + V_2) f_1^1 + \left(\frac{\mu}{2} + \partial_2^A\right) f_2^1 &=& 0, 
   \label{eq:BPSeqhss7} \\
 -(M_v - V_2) f_1^2 + \left(\frac{\mu}{2} - \partial_2^A\right) f_2^2 &=& 0.
   \label{eq:BPSeqhss8}
\end{eqnarray}
%%%%%%%%%%%%%%%%%%%%%%%%%%%%

To solve the BPS equations (\ref{eq:BPSeqhss5})-(\ref{eq:BPSeqhss8}), 
 we first have to solve the 
 constraint (\ref{consthss}) and gauge away 
 the $O(2)$ gauge degrees of freedom. 
To do that we first set the following parameterization
%%%%%%%%%%%%%%%%%%%%%%%%%%%%
\begin{eqnarray}
 \phi_1^\alpha=\frac{1}{\sqrt{2}}(f_1^{2,\alpha} + i f_2^{2,\alpha}),
 ~~~~~
 \phi_2^\alpha=\frac{1}{\sqrt{2}}(f_1^{1,\alpha} + i f_2^{1,\alpha}),
 \label{eq:repar1}
\end{eqnarray}
%%%%%%%%%%%%%%%%%%%%%%%%%%%% 
where $\alpha=1,2$, and $f_a^{i,1}=f_a^i$ and $f_a^{i,2}={\bar{f}}_a^i$.
In this basis, the action corresponds to that given by Curtright and
 Freedman~\cite{CF}. 
 The $SU(2)_R$ transformation allows us to choose 
 $\xi^{(11)}=\xi^{(22)}=0$ and $\xi^{(12)}= -i \xi$ 
($\xi \in {\bf R}, \xi >0$)  without loss of generality
\footnote{
Taking a $SU(2)_R$ transformation on the model of this case, 
we can obtain the case of $\xi^{(11)}\ne 0$ and $\xi^{(22)}\ne 0$.
}.
% \footnote{The situation corresponds to the case $b=0$ and $c \neq 0$ 
%   in (\ref{linear}) in the $N=1$ language.
%  This case can be transformed to the physically equivalent case 
%   $b \neq 0$ and $c=0$ ($\xi^{(12)}=0, \xi^{(11)}\neq 0,
%   \xi^{(22)}\neq0$) 
%   by $SU(2)_R$ transformation. 
%  This point can be clarified by relating 
%   the FI term in ${\cal N}=1$ and ${\cal N}=2$ language 
%as follows. 
%  In the ${\cal N}=2$ language, the FI term is given as 
%   ${\cal L}_{FI} = \xi_{(ij)}D_v^{(ij)}$ after $u$ integration 
%   and taking a normalisation .
%  In this form, the $SU(2)_R$ covariance is manifest.
%  Since the coefficients of the FI term $\xi^{(ij)}$
%   is $SU(2)_R$ triplet, it is represented as, for example, 
%   $\xi^{(ij)}=i (\xi^{a} \sigma^a \epsilon)^{(ij)}$ 
%   ($\xi_{(ij)} \equiv \overline{\xi^{(ij)}}=\epsilon_{ik}\epsilon_{jl}
%     \xi^{(kl)}$) where $\xi^{a}$ is a
%   real parameter, while $D_v^{(ij)}$ is defined as  
%   $D_v^{(ij)}=i (D^{a} \sigma^a \epsilon)^{(ij)}$.
%  It can be recognised as the FI term in the $N=1$ formalism by a relation
%   $\xi^1=\frac{b + \bar{b}}{2\sqrt{2}},
%    ~\xi^2=\frac{i(b-\bar{b})}{2\sqrt{2}},~\xi^3=-\frac{c}{2}$,
%    ~$D^1=\sqrt{2}{\bf Re} F,~D^2 = \sqrt{2}{\bf Im} F$ and $D^3=D$.
%  }.
It is most convenient to introduce independent fields $z^\alpha, 
\bar z^\alpha, \; \alpha =1, 2$ through 
the following Ansatz \cite{AF2,valent} 
%%%%%%%%%%%%%%%%%%%%%%%%%%%% 
\begin{eqnarray}
 \phi_1^\alpha = g(r){z^\alpha \over \sqrt{r}},
\qquad % ~~
 \phi_2^\alpha = 
 f(r)i\sigma^{2\alpha\beta}{{\bar{z}}^\beta \over \sqrt{r}}
 \label{eq:repar2},
\end{eqnarray} 
with the constraint 
\begin{eqnarray}
% \quad %{\rm det} {\phi_i^\alpha}
 \phi_1^1\phi_2^2-\phi_1^2\phi_2^1= - z^1{\bar{z}}^1-z^2{\bar{z}}^2, 
 \label{eq:phi_z}
\end{eqnarray} 
%%%%%%%%%%%%%%%%%%%%%%%%%%%%
where $z^\alpha$ are complex fields and
 $r=z^1{\bar{z}}^1+z^2{\bar{z}}^2$.
The real functions $f(r)$ and $g(r)$ 
are uniquely determined 
by the constraints (\ref{consthss}) and (\ref{eq:phi_z}) as 
%%%%%%%%%%%%%%%%%%%%%%%%%%%%
\begin{eqnarray}
f(r)^2 = %\frac{1}{r}(
-\xi + \sqrt{r^2 + \xi^2}
%)
,
\qquad  %~~
g(r)^2 = %\frac{1}{r}(
\xi + \sqrt{r^2 + \xi^2}
%)
.
\end{eqnarray}
%%%%%%%%%%%%%%%%%%%%%%%%%%%%
At this stage, the action can be described by the independent 
 complex fields $z^\alpha$.
Finally we set for later convenience
\footnote{About the domain of the coordinates, see Ref.~\cite{EH}.},
%%%%%%%%%%%%%%%%%%%%%%%%%%%%
\begin{eqnarray}
 & z^1 = \sqrt{r}\cos \frac{\Theta}{2} \exp \frac{i}{2} (\Psi + \Phi),&
         \label{eq:repar3} \\
 & z^2 = \sqrt{r}\sin \frac{\Theta}{2} \exp \frac{i}{2} (\Psi - \Phi),&
         \label{eq:repar4} \\
 & 0 \le r \le \infty,
\quad%~
0 \le \Theta \le \pi,
\quad%~
0 \le \Phi \le 2\pi,
\quad%~
0 \le \Psi \le 2\pi. & \nonumber 
\end{eqnarray}
%%%%%%%%%%%%%%%%%%%%%%%%%%%%
By using the above spherical coordinates, 
 the BPS equations are rewritten as
%%%%%%%%%%%%%%%%%%%%%%%%%%%%
\begin{eqnarray}
 0 & = &-\frac{e^{\frac{i}{2}(\Phi-\Psi)}}{2\sqrt{2}\sqrt{r^2+\xi^2}}
          \Bigg\{r^\prime ( {f}- {g}) \sin\frac{\Theta}{2}
          +\xi \cos \frac{\Theta}{2}( {f}+ {g})
          (\mu\sin{\Theta}+i \sin\Theta\Phi^\prime+\Theta^\prime)
         \nonumber \\
   &   &+~~~~~
          r( {f}- {g})\left(\mu\sin\frac{\Theta}{2}+
          i\sin\frac{\Theta}{2}\Phi^\prime-i\sin\frac{\Theta}{2}\Psi^\prime
          +\cos\frac{\Theta}{2}\Theta^\prime \right)\Bigg\}, 
         \label{BPSeqhsssp1} \\
 0 & = & \frac{e^{\frac{i}{2}(\Phi+\Psi)}}{2\sqrt{2}\sqrt{r^2+\xi^2}}
          \Bigg\{r^\prime ( {f}- {g})\cos\frac{\Theta}{2}
          -\xi \sin \frac{\Theta}{2}( {f}+ {g})
          (\mu\sin{\Theta}-i \sin\Theta\Phi^\prime+\Theta^\prime)
         \nonumber \\
   &   &-~~~~~
          r( {f}- {g})\left(\mu\cos\frac{\Theta}{2}-
          i \cos \frac{\Theta}{2}\Phi^\prime-i\cos \frac{\Theta}{2}\Psi^\prime
          +\sin\frac{\Theta}{2}\Theta^\prime \right)\Bigg\}, 
         \label{BPSeqhsssp2} \\
 0 & = &-\frac{ie^{\frac{i}{2}(\Phi-\Psi)}}{2\sqrt{2}\sqrt{r^2+\xi^2}}
          \Bigg\{r^\prime ( {f}+ {g}) \sin\frac{\Theta}{2}
          -\xi \cos \frac{\Theta}{2}( {f}- {g})
          (\mu\sin{\Theta}+i\sin\Theta\Phi^\prime+\Theta^\prime)
         \nonumber \\
   &   &+~~~~~
          r( {f}+ {g})\left(\mu\sin\frac{\Theta}{2}+i
          \sin\frac{\Theta}{2}\Phi^\prime-i\sin\frac{\Theta}{2}\Psi^\prime
          +\cos\frac{\Theta}{2}\Theta^\prime \right)\Bigg\}, 
        \label{BPSeqhsssp3} \\
 0 & = &-\frac{ie^{\frac{i}{2}(\Phi+\Psi)}}{2\sqrt{2}\sqrt{r^2+\xi^2}}
          \Bigg\{r^\prime ( {f}+ {g})\cos\frac{\Theta}{2}
          +\xi \sin \frac{\Theta}{2}( {f}- {g})
          (\mu\sin{\Theta}-i \sin\Theta\Phi^\prime+\Theta^\prime)
         \nonumber \\
   &   &-~~~~~
          r( {f}+ {g})\left(\mu\cos\frac{\Theta}{2}-
          i \cos \frac{\Theta}{2}\Phi^\prime-i\cos \frac{\Theta}{2}\Psi^\prime
          +\sin\frac{\Theta}{2}\Theta^\prime \right)\Bigg\},
         \label{BPSeqhsssp4}
\end{eqnarray} 
%%%%%%%%%%%%%%%%%%%%%%%%%%%%
where prime denotes the derivative in terms of $y$.
%, and $\tilde{f}$ and $\tilde{g}$ are real functions defined as follows
%%%%%%%%%%%%%%%%%%%%%%%%%%%%
%\begin{eqnarray}
% \tilde{f}(r)^2=-\xi + \sqrt{r^2 + \xi^2},
% ~~
% \tilde{g}(r)^2=\xi + \sqrt{r^2 + \xi^2}.
%\end{eqnarray}
%%%%%%%%%%%%%%%%%%%%%%%%%%%%
BPS wall solution should approach the supersymmetric discrete 
 vacua as $y\rightarrow \pm\infty$.
By eliminating the terms with $y$ derivative in 
 Eqs.~(\ref{BPSeqhsssp1})-(\ref{BPSeqhsssp4}), we find
 the supersymmetric vacuum condition 
%%%%%%%%%%%%%%%%%%%%%%%%%%%%
\begin{eqnarray}
 0 &=& \xi \cos\frac{\Theta}{2}( {f}+ {g})\mu \sin\Theta
 +r( {f}- {g})\mu\sin\frac{\Theta}{2}, \label{vacuaeq1} \\
 0 &=& \xi \sin\frac{\Theta}{2}( {f}+ {g})\mu \sin\Theta
 +r( {f}- {g})\mu\cos\frac{\Theta}{2}, \label{vacuaeq2} \\
 0 &=& \xi \cos\frac{\Theta}{2}( {f}- {g})\mu \sin\Theta
 -r( {f}+ {g})\mu\sin\frac{\Theta}{2}, \label{vacuaeq3} \\
0 &=& \xi \sin\frac{\Theta}{2}( {f}- {g})\mu \sin\Theta
 -r( {f}+ {g})\mu\cos\frac{\Theta}{2}. \label{vacuaeq4}
\end{eqnarray}
%%%%%%%%%%%%%%%%%%%%%%%%%%%%
It is found that there are only two vacua satisfying all these equations 
%are given by
: $(r,\Theta)=(0,0),(0,\pi)$.
Therefore we consider the domain wall solution connects these vacua,
and we can expect that $\Theta$ has nontrivial configuration. 
% and Eqs.~(\ref{BPSeqhsssp1})-(\ref{BPSeqhsssp4}) are consistent
% with a trivial configuration for all other components.
%Thus, we put $r=\Phi=\Psi=0$
% \footnote{In general, we can put $r=0$ and $\Phi=\Psi=$constant.}.
After some calculations, we can derive the four independent 
differential equations from Eqs.~(\ref{BPSeqhsssp1})-(\ref{BPSeqhsssp4}) 
in the following 
%%%%%%%%%%%%%%%%%%%%%%%%%%%%%
\begin{eqnarray}
r' &=& \mu \cos \Theta \cdot r, \qquad r\cdot \Psi' = 0, \label{fulleq1} \\
\Theta' &=& - \mu \sin \Theta, \qquad \sin\Theta \cdot \Phi' = 0. 
 \label{fulleq2}
\end{eqnarray}
%%%%%%%%%%%%%%%%%%%%%%%%%%%%%
%Substituting them into (\ref{BPSeqhsssp1})-(\ref{BPSeqhsssp4}), 
% we find only one equation
%%%%%%%%%%%%%%%%%%%%%%%%%%%%
%\begin{eqnarray}
% \mu \sin\Theta + \Theta^\prime=0. \label{eq:BPSeqhss}
%\end{eqnarray}
%%%%%%%%%%%%%%%%%%%%%%%%%%%%
%By requiring t
The boundary condition of $r=0$ at $y=-\infty$ 
dictates %we can obtain 
the solution 
\footnote{We may set to $\Psi = 0$ because of the singularity of 
coordinate $\Psi$ at $r=0$ on the target space.}
of (\ref{fulleq1}) to be 
$r=0$ and  $\Psi = 0$. On the other hand, 
nontrivial BPS solutions are obtaind from (\ref{fulleq2}) as 
%The equation (\ref{eq:BPSeqhss}) can be easily solved as
%%%%%%%%%%%%%%%%%%%%%%%%%%%%
\begin{eqnarray}
 \Theta=\arccos[\tanh\mu (y + y_0)], 
\qquad%\;\; 
\Phi = \varphi_0, 
\label{eq:solhss}
\end{eqnarray}
%%%%%%%%%%%%%%%%%%%%%%%%%%%%
where $y_0$ and $\varphi_0$ are real constants: 
$y_0$ determines the position of the domain wall 
along $y$ direction and $\varphi_0$ corresponds to the 
Nambu-Goldstone (NG) mode of $U(1)$ isometry of target space. 
%Plot of the wall solution is shown in Fig.~\ref{plothss}.

We can show that the solution
 $(\ref{eq:solhss})$ can be mapped to (\ref{GPT-sol}).
To see this, we put the following parameterization
\footnote{
%In fact, t
This parameterization has an ambiguity 
in the rotation of $X^1$, $X^2$ and $X^3$.
}
%%%%%%%%%%%%%%%%%%%%%%%%%%%%
\begin{eqnarray}
 X^1 &=& r \sin \Theta \cos \Psi, \label{repar5} \\
 X^2 &=& r \sin \Theta \sin \Psi, \label{repar6} \\ 
 X^3 &=& \sqrt{r^2 + \xi^2} \cos \Theta, \label{repar7} \\
 \varphi &=& \Phi + \Psi. \label{repar8} 
 \end{eqnarray}
%%%%%%%%%%%%%%%%%%%%%%%%%%%%
Substituting $r=\Psi=0$ and $(\ref{eq:solhss})$ into
 (\ref{repar5})-(\ref{repar8}), we obtain the following form 
%%%%%%%%%%%%%%%%%%%%%%%%%%%%
\begin{eqnarray}
 X^1 &=& X^2 = 0, \\
X^3 &=& \xi \tanh\mu(y + y_0), 
\qquad %\;\; 
\varphi = \varphi_0.
\label{townsend} 
\end{eqnarray} 
%%%%%%%%%%%%%%%%%%%%%%%%%%%%
%Putting $\xi=1$,
% it 
It is found that this solution coincides with 
%(\ref{GPT-sol}).
 that derived in Ref.~\cite{GTT1}
\footnote{
We find the {\it same} BPS wall solution as previous section 
in spite of solving the BPS equation associated with 
{\it different} $\frac{1}{2}$ SUSY condition. 
The relation between two $\frac{1}{2}$ SUSY conditions 
are discussed in Appendix A.
}.
We also show that the bosonic part of the action (\ref{bosonachss})
 corresponds to that given in Ref.~\cite{GTT1}.
By using the parameterisation 
 (\ref{eq:repar1})-(\ref{eq:repar4}) and (\ref{repar5})-(\ref{repar8}),
 the bosonic part of the action (\ref{bosonachss}) can be rewritten as 
%%%%%%%%%%%%%%%%%%%%%%%%%%%%
\begin{equation}\label{nlsm_lag}
 {\cal L} 
 = -{1\over 2}\left\{U\partial_\mu{\bf X}\cdot
 \partial^\mu {\bf X} + U^{-1}{\cal D}_\mu \varphi {\cal D}^\mu
 \varphi + \mu^2 U^{-1}\right\},
\end{equation}
%%%%%%%%%%%%%%%%%%%%%%%%%%%%
where
 ${\cal D}_\mu\varphi = \partial_\mu\varphi + {\bf A} \cdot 
 \partial_\mu{\bf X}$
 and 
%%%%%%%%%%%%%%%%%%%%%%%%%%%%%
\begin{eqnarray}
 {\mbox{\boldmath $\nabla$}} \times {\mbox{\bf A}} 
 = \mbox{\boldmath $\nabla$} U. \label{eq:monopole} 
\end{eqnarray}
%%%%%%%%%%%%%%%%%%%%%%%%%%%%%
The harmonic function $U$ can be described
%%%%%%%%%%%%%%%%%%%%%%%%%%%%%
\begin{equation}\label{harmonic_fun}
 U =  {1\over2}\left[{1\over |{\bf X}- \xi{\bf n}|} + 
 {1\over |{\bf X}+ \xi{\bf n}|}\right]\, ,
 \label{eq:harmonic_func}
\end{equation}
%%%%%%%%%%%%%%%%%%%%%%%%%%%%%
where ${\bf n}$ is a unit three vector, which is given by 
%${\bf n}=(1,0,0)$. 
 ${\bf n}=(0,0,1)$. 
${\bf A}$ is a potential whose solution is given as
%%%%%%%%%%%%%%%%%%%%%%%%%%%%%%%%%%%%
\begin{eqnarray}
 A_1 
 &=& {1 \over 2}
 %\frac{1}{4}
  \left\{
    \frac{X^2} 
    {|{\bf X} - \xi {\bf n}| (X^3 - \xi + |{\bf X} - \xi {\bf n}|)}
   +\frac{X^2} 
   {|{\bf X} + \xi {\bf n}| (X^3 + \xi + |{\bf X} - \xi {\bf n}|)}
   \right\}, \\
 A_2 
 &=& {1 \over 2}
 %\frac{1}{4} 
 \left\{
    \frac{-X^1} 
    {|{\bf X} - \xi {\bf n}| (X^3 - \xi + |{\bf X} - \xi {\bf n}|)}
   +\frac{-X^1} 
   {|{\bf X} + \xi {\bf n}| (X^3 + \xi + |{\bf X} + \xi {\bf n}|)}
   \right\}, \\  
 A_3 &=& 0.
\end{eqnarray}
%%%%%%%%%%%%%%%%%%%%%%%%%%%%
It is found that the target metric of the action (\ref{nlsm_lag}) 
is just the Eguchi-Hanson 
 metric \cite{CF,ivanov-EH,valent}.

Finally we give the BPS solution in terms of harmonic 
 superfields $q_a^+$. 
This is a starting point to derive the 
 LEEA
 %low energy effective action (LEEA) 
around the wall background.
They are derived by using the change of variables 
 (\ref{eq:repar1})-(\ref{eq:repar4}). 
The results are
%%%%%%%%%%%%%%%%%%%%%%%%%%%%
\begin{eqnarray}
 q_1^+=f^i_1 u_i^+ &=&\sqrt{\frac{\xi}{2}}e^{\frac{i}{2}\varphi_0}
                       \left(\begin{array}{c}
                           -\sqrt{1-\tanh(\mu (y+y_0))} u_1^+ \\     
                            \sqrt{1+\tanh(\mu (y+y_0))} u_2^+  
                       \end{array}
                      \right), \\
 q_2^+=f^i_2 u_i^+ &=&-i\sqrt{\frac{\xi}{2}}e^{\frac{i}{2}\varphi_0}
                      \left(\begin{array}{c}
                            \sqrt{1-\tanh(\mu (y+y_0))} u_1^+ \\     
                            \sqrt{1+\tanh(\mu (y+y_0))} u_2^+  
			    \end{array}
                      \right). 
\end{eqnarray}
%%%%%%%%%%%%%%%%%%%%%%%%%%%%
%%%%%%%%%%%%%%%%%%%%%%
%                    %
% Section 4          %
%                    %
%%%%%%%%%%%%%%%%%%%%%%
%%%%%%%%%
\section{Discussion
}\label{sc:discussion}
%%%%%%%%%%%%%%%%%%%%%%%%%%%%%%%%%%%%%%

Obtaining the LEEA
 %low energy effective action (LEEA) 
 is usually 
one of the objectives to study models with domain walls. 
There have been a number of works studying linear sigma models. 
Since nonlinear sigma model is often necessary to consider 
eight or more SUSY, 
we shall consider LEEA for nonlinear sigma model 
in this section. 
To illustrate an issue, we use component formalism here and leave 
the treatment in the superfield formalism for future work. 

Let us take the component action of the $T^*{\bf C}P^1$ model 
in (\ref{nlsm_lag}).
%For simplicity, we put $\xi=1$.
In order to obtain a 
 ${1 \over 2}$ BPS wall solution, it has been found that one can 
 consistently truncate \cite{GTT1} 
 by taking $X^1, X^2=0$ to obtain a truncated model 
 with $\varphi$ and $X \equiv X^3$ 
 whose range is $|X|\le \xi$ with 
 $U=\xi/(\xi^2-X^2)$ 
%%%%%%%%%%%%%%%%%%%%%%%%%%%
\begin{equation}\label{truncated_nlsm}
\int d^4x {\cal L}_{\rm truncated} 
= -{1\over 2}\int d^4x  \left\{U\partial_\mu X 
\partial^\mu  X + U^{-1} \partial_\mu \varphi \partial^\mu \varphi
+ \mu^2 U^{-1}\right\}.
\end{equation}
If we assume the field configurations to depend on a single coordinate y,
energy density in the truncated model can be rewritten as 
\begin{eqnarray}
  {\cal E}_{\rm truncated} 
   &=& \frac{1}{2}\left[U\left( \partial_2X \right)^2
       + U^{-1}\left( \partial_2\varphi\right)^2 +\mu^2 U^{-1}\right] 
       \nonumber \\
   &=& \frac{1}{2}\left[U(\partial_2 X - \mu U^{-1})^2 
   + U^{-1}(\partial_2\varphi)^2\right]
       +\partial_2\left(\mu X\right), 
\end{eqnarray}
and one can obtain the ${1 \over 2}$ BPS equation as 
\begin{eqnarray}
  \frac{\partial \varphi}{\partial y}=0, \qquad 
  \frac{\partial X}{\partial y}= \mu U^{-1} = \mu {\xi^2-X^2 \over \xi}.
\label{BPS-eq}
\end{eqnarray}
The BPS wall solutions become 
\begin{eqnarray}
  X_{\rm cl} = \xi \tanh \mu(y+y_0), \qquad 
  \varphi_{\rm cl} = {\rm constant} = \varphi_0 .
\label{BPS-sol}
\end{eqnarray}
Without loss of generality, we can choose $y_0=0$ and $\varphi_0=0$.

To describe the field theory of fluctuations on the background, 
we decompose the fields in terms of mode functions which 
are usually defined by %linearised 
 linearized equations of motion 
\begin{equation}
X(x, y) = X_{\rm cl}(y)+ \sum_n X_n(x) a_n(y), 
\qquad  
\varphi(x, y) =  \varphi_{\rm cl} (y) + 
 \sum_n \varphi_n(x) b_n(y)\,, 
\end{equation}
where $x$ denotes three-dimensional world volume coordinates of of 
domain wall.
Among various bosonic modes, 
one can easily find massless modes corresponding to 
spontaneously broken global symmetry generators, namely 
the Nambu-Goldstone (NG) particles corresponding to 
translation and the $U(1)$ isometry 
\begin{equation}
a_0(y) \equiv {d X_{\rm cl} \over dy} 
= {\mu\xi \over \cosh \mu y}, 
\qquad  
b_0(y) \equiv  
\left.{d \varphi_{\rm cl} (y) \over d \varphi_0}\right|_{\varphi_0=0} 
=1. 
\label{eq:zero_mode_wf}
\end{equation}

We are usually interested in LEEA of 
massless or nearly massless modes. 
If decoupling holds, massive modes give contributions suppressed 
by inverse powers of 
their masses after functional integration. 
In such a circumstance, 
we can obtain LEEA by retaining only the 
massless modes 
$X(x, y) \rightarrow X_{\rm cl}(y)+ X_0(x) a_0(y)$, and 
$\varphi(x, y) =  \varphi_{\rm cl} (y) + 
 \varphi_0(x) b_0(y)$.
After keeping only zero modes $X_0(x), \varphi_0(x)$ 
and integrating over $y$, 
we obtain a candidate of LEEA %the low energy effective Lagrangian 
%to all powers of fields $X_0, \varphi_0$ 
\begin{equation}\label{eff_lag_X}
\int d^3x {\cal L}_{\rm LE \: eff} =
\int d^3x \left({\cal L}_{\rm kin \varphi} 
+
{\cal L}_{\rm kin X}
+
{\cal L}_{\rm pot}
+{\cal E}_{\rm bkgr} \right),
\end{equation}
\begin{equation}
{\cal L}_{\rm kin \varphi} =
 -{\xi \over 2} \left({2 \over \mu}-{4 \over 3}\mu X_0^2\right)
(\partial_a \varphi_0)^2, 
%\end{equation}
%\begin{equation}
\quad 
{\cal L}_{\rm kin X}
= {1\over 2} {\xi \over 2\mu X_0^2} \log\left(1-4 \mu^2 X_0^2\right)
(\partial_a X_0)^2, 
\end{equation}
\begin{equation}
{\cal L}_{\rm pot}
= -{\xi\mu \over 2} 
 \left\{6 -{4 \over 3}\mu^2 X_0^2 
 +{2 \over \mu X_0}\log \left({1-2\mu X_0 \over 1+2\mu X_0}\right)
 -\left(2+{1 \over 2\mu^2 X_0^2} \right)
 \log \left(1-4\mu^2X_0^2\right)\right\},
 \label{eq:eff_lag_pot}
\end{equation}
where index $a=0,1,3$ denotes spacetime dimensions corresponding 
to the world volume of the wall, and ${\cal E}_{\rm bkgr}=2\mu\xi$ 
is the energy density of the wall. 
Let us note that 
the $U(1)$ zero mode $\varphi_0$ is normalizable 
in spite of the constant wave function $b_0(y) \equiv 1$ 
in Eq.(\ref{eq:zero_mode_wf}). 
This shows that the $U(1)$ zero mode is effectively localized 
because of the nonlinearity of the kinetic term, 
whereas the wave function $a_0(y)$ 
of the translation zero mode $X_0$ 
is manifestly localized. 
By examining the linearized equation of motion for the 
three-dimensional fields $X_0(x)$ and $\varphi_0(x)$, 
we find that the masses of these fields vanish. 
On the other hand, there are nonlinear interaction terms 
both of derivative and non-derivative types in 
Eqs.~(\ref{eff_lag_X})-(\ref{eq:eff_lag_pot}). 
There are two problems leading to the above result. 
One problem is that decoupling of massive modes requires 
to retain auxiliary fields of the remaining SUSY \cite{Sakamura}. 
The other problem is the possible field redefinition 
which we now wish to examine. 
%: there may be a better choice of fields than $X_0, \varphi_0$. 

In choosing a field variable for zero modes, we can take the 
celebrated Manton's method as a guiding principle. 
He has proposed to obtain 
 (a part of) LEEA 
%the effective action 
of zero modes \cite{Manton}, \cite{To}. 
When a soliton has a moduli such as a translational collective 
coordinate, 
one promotes the collective coordinate to a field on the soliton 
world volume. 
Since the method presupposes a slow motion in the moduli 
space, the Manton's method should give %the effective action 
LEEA of the zero mode at least up to two derivative terms 
correctly. 
The collective coordinates $y_0$ for the translation and $\varphi_0$ 
for the $U(1)$ isometry appear in the classical solution as 
in Eq.~(\ref{BPS-sol}). 
Promoting the collective coordinates to the zero mode fields 
$y_0(x), \varphi_0(x)$ by the Manton's method, we obtain 
\begin{eqnarray}
\label{eq:transl_zero_mode}
X(x, y)
&\!\!\! 
=&\!\!\! 
\xi \tanh \mu (y+y_0(x))=\left. X_{\rm cl}(y)\right|_{y_0=0}
+\left. {dX_{\rm cl} \over dy}(y)\right|_{y_0=0} y_0+ O(y_0^2) + \cdots , \\
%\qquad  
\varphi (x, y) 
&\!\!\! 
=
&\!\!\! 
 \varphi_0(x) 
=\left. {d \varphi_{\rm cl} \over dy}(y)\right|_{\varphi_0=0}  \varphi_0(x). 
\end{eqnarray}
As we have seen, the zero mode wave function 
is usually obtained by a derivative of the classical field configuration. 
This usual definition of zero mode field coincides with the 
Manton's method only when we choose the field variables as 
 $u, \varphi$ 
by a field redefinition $X \rightarrow u$ 
\begin{equation}
X 
= \xi \tanh u, \\
\qquad 
v \equiv e^{\Omega} \equiv e^{u + i \varphi}.
\end{equation}
We can identify $u, \varphi$ as real and imaginary part 
of a chiral scalar field $\Omega$, 
since a scalar multiplet of ${\cal N}=2$ SUSY in three dimensions 
requires a complex scalar field.
With $u, \varphi$, the truncated model is given by a massive ${\bf C}P^1$ 
nonlinear sigma model 
\begin{eqnarray}
\int d^4x {\cal L}_{\rm truncated} 
&
=
&
 -{\xi\over 2}\int d^4x  {1 \over \cosh^2 u} 
 \left((\partial_\mu u)^2 + 
 (\partial_\mu \varphi)^2 + \mu^2 \right) \nonumber \\
&
=
&
 -{\xi\over 2}\int d^4x  {4 \over (1+|v|^2)^2 } 
 \left(\partial_\mu v^* \partial^\mu v 
  + \mu^2 v^* v \right)
\end{eqnarray}
where the Fubini-Study metric of ${\bf C}P^1$ can be recognised. 
This choice of field precisely corresponds to the truncated model 
with the ${\cal N}=1$ superfield Eqs.~(\ref{CP1}) and (\ref{CP1V}). 
BPS wall solution and the NG mode function for translation is 
given by 
\begin{equation}
u_{\rm cl}(y)= \mu (y + y_0), 
\qquad 
a_{u 0}(y)\equiv {d u_{\rm cl}(y) \over dy}=\mu.
\end{equation}

After decomposing into modes $u(x, y)= u_{\rm cl}(y)+\sum u_n(x) a_{u n}(y)$, 
retaining fields $u_0(x), \varphi_0(x)$ corresponding 
to zero modes and integrating over $y$, we find precisely a free massless 
complex scalar action 
\begin{eqnarray}\label{eq:LEEA_u}
\int d^3x {\cal L}_{\rm LE \: eff} 
&
=
&
 -{\mu\xi\over 2}\int d^3x   
 \left((\partial_a u_0)^2 + 
 (\partial_a \varphi_0)^2 \right) 
 + {\cal E}_{\rm bkgr}.
\end{eqnarray}
We have computed all powers in the fields $u_0, \varphi_0$ 
without making approximations. 
Let us observe that the field redefinition from $X_0(x)$ to $u_0(x)$ 
is not just an ordinary field redefinition local in $x$. 
It involves all higher massive modes and functions of $y$ 
in a complicated way 
as one can see by comparing mode expansions 
\begin{equation}
X(x,y)=X_{\rm cl}(y)+\sum_n X_n(x) a_n(y)
=\xi\tanh u(x, y) 
=\xi \tanh \left(u_{\rm cl}(y)+ \sum_n u_n(x) a_{u n} (y) \right).
\end{equation}
We have no reason to believe that the LEEA in 
Eq.~(\ref{eq:LEEA_u}) can be obtained from Eq.~(\ref{eff_lag_X}) 
by a local field redefinition $u_0(x)=f\left(X_0(x)\right)$. 
Therefore it is important to choose the appropriate variable to define 
zero mode field. 
With the Manton's method, 
we are at least sure that the zero modes $u_0(x), \varphi_0(x)$ 
have no interactions as far as two derivative terms are concerned 
in conformity with the low energy theorems. 
LEEA of NG particles associated with walls can be obtained also 
by means of nonlinear realization \cite{CNV}, 
whose result is consistent with ours up to two derivatives as expected. 

Let us also note that the choice of $u, \varphi$ has also allowed the 
identification of these fields as a complex scalar field of the 
chiral scalar multiplet of ${\cal N}=2$ SUSY in three dimensions 
($4$ supercharges). 
Thus the choice of the field variable is better for preserved SUSY 
which requires the two real scalar fields for zero mode 
to have the identical wave function. 

When one wants to obtain the LEEA, 
 K\"ahler Normal Coordinate (KNC) \cite{HIN} may be useful 
to represent all modes. 
Using the KNC, any tensor can be expanded covariantly to keep 
 the complex structure.
Then, the general action described by K\"ahler potential and superpotential 
 can be expanded, and on-shell action is obtained in terms of 
 boson and fermion fields, which is described by the KNC geometrically. 
Applying the KNC with the real domain wall solution as the background and 
 expanding the general action, it is easy to find the 
 correspondence between mode equations for boson 
 and fermion fields in all modes including massive modes.
In Appendix \ref{KNC}, we show such a correspondence.   

\vspace{1.0cm}

\noindent{\Large \bf Acknowledgements}\\

\noindent 
We are grateful for discussion with Sergei Ketov. 
 One of the authors (NS) thanks useful discussion with Taichiro Kugo. 
This work is supported in part by Grant-in-Aid 
 for Scientific Research from the Japan Ministry 
 of Education, Science and Culture  13640269. 
The work of M.~Naganuma is supported by JSPS Fellowship. 
The work of M.~Nitta is supported by the U.~S. Department
 of Energy under grant DE-FG02-91ER40681 (Task B).

%%%%%%%%%%%%%%%%%%%%%%%
\renewcommand{\thesubsection}{\thesection.\arabic{subsection}}

\appendix

%%%%%%%%%%%%%%%%%%%%%%%%%%%%%%%%%%%%%%%%%%%%%%%%%%%%%%%%%%%%
%\section{$SU(2)_R$, $1/2$ SUSY and Lorentz invariance}
\section{Two $1/2$ SUSY conditions and Lorentz invariance}
%%%%%%%%%%%%%%%%%%%%%%%%%%%%%%%%%%%%%%%%%%%%%%%%%%%%%%%%%%%%
%%%%%%%%%%%%%%%%%%%%%%%%%%%%%%%%
\subsection{$1/2$ SUSY condition}\label{half-SUSY}
%%%%%%%%%%%%%%%%%%%%%%%%%%%%%%%%
In this Appendix, we derive the half SUSY condition (\ref{eq:half-SUSY}), 
%which preserves four SUSY charges. 
and find a $SU(2)_R$ transformation between the two models 
discribed in ${\cal N}=1$ superfield and the HSF, 
by examining the relation of $\frac{1}{2}$ SUSY conditions. 
 We also show that models allowed by this condition are Lorentz invariant 
only for three dimensional spacetime.

In order to find SUSY conserved by BPS wall background,
let us examine SUSY transformation of fermions in bosonic background.
The SUSY transformation in ${\cal N}=2$ SUSY massive 
 nonlinear sigma model in four dimensions ($D=4$) is obtained from 
 the SUSY transformation in ${\cal N}=1$ SUSY nonlinear sigma models 
in six dimensions ($D=6$) with nontrivial dimensional reduction. 
The SUSY transformation of fermions in $D=6$ is given by 
\begin{eqnarray}
  \delta \chi^a = \Gamma^M f^{ai}_X \partial_M \phi^X \varepsilon_i, \qquad 
%\;\;
  (X=1,..,4n),
\label{SUSY-transf}
\end{eqnarray}
where $\Gamma^M$$(M=0,1,...,5)$ are the $D=6$ Dirac matrices, 
$f^{ai}_X$$(a=1,..,2n,i=1,2)$ are fielbeins of target 
hyper-K\"ahler manifold, transforming in the $(2n,2)$ representation 
of $Sp_n \times Sp_1$, and $\varepsilon_i$ is a $Sp_1$-Majorana and 
Weyl spinor satisfying $\Gamma^{012345}\varepsilon^i = \varepsilon^i$.
On-shell SUSY transformation of fermions in $D=4$ ${\cal N}=2$ 
SUSY nonlinear sigma models with potential terms can be obtained 
by the nontrivial dimensional reduction from $D=6$ to $D=4$ \cite{GTT1}
\begin{eqnarray}
 \frac{\partial \phi^X}{\partial x^4} = 0, \qquad 
 \frac{\partial \phi^X}{\partial x^5} = \mu k^X,
 \label{SS-red}
\end{eqnarray}
where $k\equiv k^X\partial_X$ is a tri-holomorphic Killing vector, 
 which is the same as Eq.~(\ref{eq:killing}), and 
 $\mu$ is a real mass parameter.

Substituting the Eguchi-Hanson metric and corresponding Killing vector 
to the r.h.s. of Eqs.~(\ref{SUSY-transf}) and (\ref{SS-red}), 
and requiring the vanishing SUSY 
variation, we obtain the condition of SUSY configuration, 
after some algebra \cite{SierraTownsend,GTT1},
\begin{eqnarray} 
  \Gamma^\mu[{\mbox{\boldmath $\rho$}}_i{}^j\cdot \partial_\mu {\bf X} 
  +\delta_i{}^j\cdot iU^{-1}{\cal D}_\mu\varphi]
  \varepsilon_j= -i \mu U^{-1} \Gamma^5 \varepsilon_i, \qquad 
%\;\; 
(\mu=0,...,3).
\label{SUSY-cond}
\end{eqnarray}
When we substitute the BPS equations (\ref{BPS-eq}) into the 
Eq.(\ref{SUSY-cond}), 
we obtain the $\frac{1}{2}$ BPS condition for wall solution as
%%%%%%%%%%%%%%%%%
\begin{eqnarray}
  \Gamma^{25}(\rho^3){}_i{}^j\varepsilon_j = -i \varepsilon_i.
\label{BPS-COND}
\end{eqnarray}
%%%%%%%%%%%%%%%%%
If we choose the chiral representation of Dirac matrices such as 
%%%%%%%%%%%%%%%%
\begin{eqnarray}
  \Gamma^\mu = \gamma^\mu\otimes\tau^1, \qquad 
%\; 
\Gamma^4=\gamma^5\otimes \tau^1, \qquad 
%\;
  \Gamma^5 = i{\bf 1}_4\otimes \tau^2,
\end{eqnarray}
%%%%%%%%%%%%%%%
then spinor parameters can be reduced four component spinors 
like $(\varepsilon_i)^T = ((\varepsilon_i)^T,0,0,0,0)$, 
corresponding to two sets of Majorana spinors in $D=4$.
Moreover the Eq.~(\ref{BPS-COND}) can be rewritten as
%%%%%%%%%%%%%%%
\begin{eqnarray} 
  \sigma^2 \bar{\epsilon}^1=i\epsilon_1, \qquad 
  \sigma^2 \bar{\epsilon}^2=-i\epsilon_2,
\label{BPS-cond}
\end{eqnarray}
%%%%%%%%%%%%%%%
where $\epsilon_i$ are two Weyl spinors in $D=4$, such as 
$(\varepsilon_i)^T \equiv ((\epsilon_i)^T, (\bar{\epsilon}^i)^T)$.
%%%%%%%%%%%%%%%
%\begin{eqnarray}
%
%\begin{eqnarray}
%%%%%%%%%%%%%%
This condition coincides with (\ref{N1harfSUSY}), 
if we identity $\epsilon = \epsilon_1$, 
for $\alpha = 0$ corresponding to the case of 
real parameters $b$ and $\mu$.
From the Eq.~(\ref{BPS-cond}) we can see that BPS wall solution 
conserves four SUSY out of eight SUSY in four dimensions.

%%%%%%%%%%%%%%%%%%%%%%%%%%%%%%%%
\subsection{$SU(2)_R$ transformation}\label{SU2R}
%%%%%%%%%%%%%%%%%%%%%%%%%%%%%%%%
We show that $\frac{1}{2}$ SUSY condtion 
(\ref{eq:half-SUSY}) is related with (\ref{BPS-cond}) 
by an $SU(2)_R$ transformation. 
Let us take a $SU(2)$ transformation generated by 
%%%%%%%%%%%%%
\begin{eqnarray}
g_i{}^j \equiv \frac{1}{\sqrt{2}}\left(
\begin{array}{cc}
1 & -1 \\
1 & 1 
\end{array}\right)
\label{su2rtrans}
\end{eqnarray}
%%%%%%%%%%%%%
on the both sides of Eq.(\ref{BPS-COND}). 
Then we obtain the new condition 
%If one truncate by taking $X^2=X^3=0$ and $X=X^1$, 
% the following BPS condition can be obtained
%%%%%%%%%%%%%%%%%
\begin{eqnarray}
 \Gamma^{25}(\rho^1)_i{}^j \varepsilon'_j=-i \varepsilon'_i.
\label{BPS-COND2}
\end{eqnarray}
%%%%%%%%%%%%%%%%
given by new basis of spinor parameters 
$\varepsilon'\equiv g \varepsilon$.
This can be rewritten, in two Weyl spinors in $D=4$, as
%%%%%%%%%%%%%%%%
\begin{eqnarray}
\sigma^2 \bar{\epsilon'}^1=i \epsilon'_2=i \epsilon'^1,
~~~~~\sigma^2 \bar{\epsilon'}^2=i \epsilon'_1=-i \epsilon'^2,
\label{BPS-cond2}
\end{eqnarray}
%%%%%%%%%%%%%%%%
where $(\varepsilon'_i)^T \equiv ((\epsilon'_i)^T, (\bar{\epsilon}'^i)^T)$.
This new condition coincides with (\ref{eq:half-SUSY}) and 
was used in subsection \ref{BPShss} to derive BPS equations.
Therefore we can see that two $\frac{1}{2}$ SUSY conditions 
in this paper are related with each other by the 
$SU(2)_R$ transformation given by (\ref{su2rtrans}),
which also relates between two models discussed in ${\cal N}=1$ 
superfield and in HSF.

A possible identification of 
   the FI term in ${\cal N}=1$ language and ${\cal N}=2$ (HSF) language 
is given as follows. 
  In the ${\cal N}=2$ language, the FI term is given as 
  ${\cal L}_{FI} = -\xi_{(ij)}D_v^{(ij)}/3$
% ${\cal L}_{FI} = \xi_{(ij)}D_v^{(ij)}$ 
  after $u$ integration 
   and taking a normalisation .
  In this form, the $SU(2)_R$ covariance is manifest.
  Since the coefficients of the FI term $\xi^{(ij)}$
   is $SU(2)_R$ triplet, it is represented as, for example, 
   $\xi^{(ij)}=i \xi^{a} \epsilon^{ik}\sigma^a{}_k{^j}$ 
%  $\xi^{(ij)}=i (\xi^{a} \sigma^a \epsilon)^{(ij)}$ 
   ($\xi_{(ij)} \equiv \overline{\xi^{(ij)}}=\epsilon_{ik}\epsilon_{jl}
     \xi^{(kl)}$) where $\xi^{a}$ is a
   real parameter, while $D_v^{(ij)}$ is defined as  
   $D_v^{(ij)}=i D^{a}\epsilon^{ik} \sigma^a{}_k{}^j$.
%   $D_v^{(ij)}=i (D^{a} \sigma^a \epsilon)^{(ij)}$.
  It can be recognised as the FI term in the ${\cal N}=1$ formalism 
   by a relation 
   $\xi^1=-{\rm Re}b,~\xi^2={\rm Im}b,~\xi^3=-\frac{c}{2}$, 
   $D^1 = -3 {\rm Re} F,~D^2 = -3 {\rm Im} F$
   and $D^3=-\frac{3}{2} D$.
%$\xi^1=\frac{b + \bar{b}}{2\sqrt{2}},
%    ~\xi^2=\frac{i(b-\bar{b})}{2\sqrt{2}},~\xi^3=-\frac{c}{2}$,
%    ~$D^1=\sqrt{2}{\bf Re} F,~D^2 = \sqrt{2}{\bf Im} F$ and $D^3=D$.

%%%%%%%%%%%%%%%%%%%%%%%%%%%%%%%%
\subsection{Lorentz symmetry}\label{lorentz}
%%%%%%%%%%%%%%%%%%%%%%%%%%%%%%%%
From (\ref{BPS-cond}), 
Majorana spinor parameters $\varepsilon_{\| i}$ ($\varepsilon_{\bot i}$) 
of SUSY conserved (broken) by the wall can be rewritten, 
by using projection operator, as
\begin{eqnarray}  
  {\varepsilon_\|}_i &\equiv &{\cal P}\varepsilon_i \equiv 
  \frac{1}{2}\left( 1_4 -i\gamma^2 \sigma^3{}_i{}^j\right)\varepsilon_j,
\qquad %\;\;
  {\cal P}^\dagger = {\cal P},
\label{1/2unbroken} \\
{\varepsilon_\bot}_i &\equiv &(1_4-{\cal P})\varepsilon_i \equiv 
  \frac{1}{2}\left( 1_4 +i\gamma^2 \sigma^3{}_i{}^j\right)\varepsilon_j. 
\label{1/2broken}
\end{eqnarray}
Supercharges corresponding to conserved (broken) $\frac{1}{2}$ SUSY are 
given as, in Majorana representation,  
\begin{eqnarray}
  \bar{\varepsilon}^i Q_i \equiv 
       \bar{\varepsilon}_\|^i Q_{\| i}
%       \bar{\tilde{\epsilon}}_i \tilde{Q}^i 
     + \bar{\varepsilon}_\bot^i Q_{\bot i},\quad 
  Q_{\| i}  \equiv (1_4-{\cal P})Q_i, \quad
  Q_{\bot i} \equiv {\cal P}Q_i, 
\label{1/2charge}
\end{eqnarray}
where the bar denotes the Dirac conjugate 
 $\bar{\varepsilon}=\varepsilon^\dagger \gamma_0$.
Original ${\cal N}=2$ SUSY algebra in four dimensions is 
\begin{eqnarray}
  \left[\bar{\varepsilon}^i Q_i, \bar{\eta}^j Q_j\right] 
     &=&(\varepsilon_i)^\dagger (\gamma_0) \,\{Q_i, (Q_j)^\dagger 
     (\gamma_0)\}\,\eta_j 
  \nonumber \\
   &=&(\varepsilon_i)^\dagger (\gamma_0)
       \bigg[2\gamma^\mu P_\mu\delta_i^j
       +\{1_4\Re(Z_{ij})-\gamma^5\Im(Z_{ij})\}\bigg]\eta_j.   
\label{SUSY-alg}
\end{eqnarray}  
From Eqs.~(\ref{1/2unbroken}) and (\ref{1/2charge}), the algebra of 
$\frac{1}{2}$ SUSY 
(4 SUSY) conserved by the wall becomes 
\begin{eqnarray}
  \left[\bar{\varepsilon}^i Q_{\| i}, \bar{\eta}^j Q_{\| j}\right] 
     \!\!&=&\!\!(\varepsilon_{\| i})^\dagger (\gamma_0) \,
        \{Q_i, (Q_j)^\dagger (\gamma_0)\}\,\eta_{\| j} \nonumber \\
     \!\!&=&\!\! (\varepsilon_i)^\dagger (\gamma_0) (1_4-{\cal P})\,
        \{Q_i, (Q_j)^\dagger (\gamma_0)\}\,{\cal P}\eta_j \nonumber \\ 
  \!\!&=&\!\!(\varepsilon_i)^\dagger (\gamma_0)\,
       \bigg[2\gamma^a P_a\delta_i^j+\{1_4\Re(Z_{ij})\}\bigg]\,
       {\cal P}\eta_j,\;\;
     (a=0,1,3).   
\label{1/2SUSY-alg}
\end{eqnarray}
We find in Eq.~(\ref{1/2SUSY-alg}) that 
 the momentum along $y$ direction 
does not appear in 
the r.h.s. of the commutators of four supercharges conserved by the wall, 
but it appears in the r.h.s. of the commutators of supercharges 
broken and unbroken by the wall. 
The part (\ref{SUSY-alg}) of algebra of SUSY charges 
in four dimensions conserved by the wall 
 is equivalent to the ${\cal N}=2$ 
SUSY algebra in three dimensions.    
Therefore the theory which maintains four supercharges conserved by the wall 
has Lorentz invariance of three dimensions, but not of four dimensions. 
We have thus understood this result, as due to the fact that 
$\frac{1}{2}$ SUSY condition of the wall breaks the four-dimensional 
Lorentz invariance preserving only three dimensional one.  
%%%%%%%%%%%%%%%%%%%%%%%%%%%%%%%%%%%%%%%%%%%%%%%%%%
%
% Hitchin part
%
%%%%%%%%%%%%%%%%%%%%%%%%%%%%%%%%%%%%%%%%%%%%%%%%%%
%%%%%%%%%%%%%%%%%%%%%%%%%%%%%%%%%%%%%%%%%%%%%%%%%
%\include{hitchin2}
%%%%%%%%%%%%%%%%%%%%%%%%%%%
%% Hitchin coordinates
%%%%%%%%%%%%%%%%%%%%%%%%%%%
%\documentclass[12pt]{article}

%\usepackage{amsmath,amssymb}

%\makeatletter
%\renewcommand{\theequation}{%
%   \thesection.\arabic{equation}}
% \@addtoreset{equation}{section}
%\makeatother
%%
%
\section{Hitchin coordinates} \label{Hitchin}
\newcommand{\vX}{{\bf X}}

In this appendix, we give the ${\cal N}=1$ superfield formulation 
of the massive HK sigma model on 
the asymptotically localy Euclidean (ALE) space, 
which inculdes $T^*{\bf C}P^1$. 
Then, we obtain the BPS wall solution in the case of $T^*{\bf C}P^1$. 
The multi-center metric of ALE may be used for 
 the construction of an intersecting lumps solution~\cite{NNS1}
 or a domain wall junction~\cite{NNS2}.

Let $\ph$ and ${\bf X}=(X^1,X^2,X^2)$ be coordinates 
of a four dimensional HK manifold with 
an $U(1)$ isometry of a constant shift of $\ph$. 
The metric defined by \cite{GH} 
\beq
 && d s^2 = U (\vX) d \vX \cdot d \vX 
  + U^{-1} (\vX) (d \ph + {\bf A} \cdot d \vX )^2 \;, \\
 && \nabla \times {\bf A} = \nabla U \;, 
\eeq
is the multi-center ALE space if 
$U$ is given by 
\beq
 U = \1{2} \sum_{i=1}^k \1{|\vX - \vX_i|} \;. 
\eeq
Here, $k$ is the number of the centers, 
$\vX_i$ is the position of the $i$-th center in 
the three dimensional space of $\vX$, 
and ${\bf A} = (A_1,A_2,A_3)$ is a potential.  
The case of $k=2$ corresponds to the Eguchi-Hanson 
space, $T^*{\bf C}P^1$.

We introduce complex coordinates using the method 
of Hitchin~\cite{Hi}. Here we follow  Ref.~\cite{NOY}. 
 For simplicity we take the parameter $\xi=1$ in 
 Eq.(\ref{eq:harmonic_func}) here. 
 First, we take a gauge of 
\beq
 && A_1 = 0 \;, \non
 && A_2 
 = \1{2} \sum_i {X^3 - X^3_i \over 
     |\vX - \vX_i| (X^1 - X^1_i - |\vX - \vX_i|)} \;, \non
 && A_3 
 = \1{2} \sum_i {- (X^2 - X^2_i) \over 
     |\vX - \vX_i| (X^1 - X^1_i - |\vX - \vX_i|)} \;.
\eeq
Defining the complex coordinates $\ph^i = (v,w)$ by 
\beq
 && v = X^2 + i X^3 \;, \\
 && w = C e^{- i\ph} \prod_{i=1}^k 
 \sqrt {-b + b_i + \Delta_i} \;. 
\eeq
the metric becomes 
\beq
 d s^2 = U |dv|^2 
+ U^{-1} \left| {dw \over w} - \delta d v\right|^2 \;. 
 \label{metric-1}
\eeq
Here, we have defined 
\beq
 \Delta_i \equiv |\vX - \vX_i| \;,\hs{5}
 \delta \equiv \1{2} \sum_i 
 {(b - b_i) + \Delta_i \over \Delta_i (v - e_i)},
\eeq
and 
\beq
 b \equiv X^1 \;, \hs{5} b_i \equiv X^1_i \;,\hs{5}
 e_i \equiv X^2_i + i X^3_i \;.
\eeq
In the derivation of (\ref{metric-1}), we have used 
\beq
 {d w \over w} = - U d b - i d \ph + {\rm Re}(\delta d v) \;.
 \label{dy}
\eeq 
The components of the metric and its inverse 
in the coordinates $\ph^i = (v,w)$ are obtained as
\beq
&& g_{ij^*} =
 \left(\begin{array}{cc} 
   U + U^{-1}|\delta|^2 & -U^{-1} {\delta\over w^*} \\
  -U^{-1} {\delta^* \over w} & U^{-1} {1 \over |w|^2}
 \end{array}\right) \;, \\
&& g^{ij^*} =
 \left(\begin{array}{cc} 
  U^{-1} & U^{-1} {\delta^* w^* } \cr
  U^{-1} {\delta w } 
   & {|w|^2 } (U + U^{-1}|\delta|^2)
 \end{array}\right) \;, 
\eeq
respectively. 

Since we know already that the scalar potential is given 
[see Eq.~(\ref{nlsm_lag})] by 
\beq
 V = \mu^2 U^{-1} \;, 
\eeq
we find the superpotential 
\beq
 W = \mu v \;. 
\eeq
There are $k$ isolated vacua given by $\vX = \vX_i$.

The BPS equation, $\del_2 \ph^i=-e^{i\alpha}g^{ij^*}\del_{j^*} W^*$, 
for BPS walls in the general ALE space becomes 
\beq
 \del_2 v = e^{i\alpha}\mu U^{-1} \;,\hs{5} 
 \del_2 w = e^{i\alpha}\mu \delta w U^{-1} 
 \label{BPSeq}\;.
\eeq
~From now on, 
we concentrate on the Eguchi-Hanson space (two centers).
Set the two centers as 
\beq
 && (b_1, e_1) = (0,i), \; \hs{5} (b_2, e_2) = (0,-i) \;,
\eeq
Let us set $X^1=X^2=0$ and define $v = i X^3 \equiv i X$.   
Then, 
\beq
 U = {1 \over 1 - X^2} \;, \hs{5} 
 \delta = i U ,
\eeq
and the BPS equations (\ref{BPSeq}) become
\beq
 i \del_2 X = e^{i\alpha} \mu (1- X^2) \;,\hs{5}
 \del_2 w = i e^{i\alpha} \mu w \;.
\eeq
Now we must choose $\alpha = \frac{\pi}{2}$, and then 
we see that $|w|=0$ and $\arg(w)=\mbox{const.}$ satisfy 
the second equation.
The BPS domain wall solution is thus obtained as
\beq
 -iv = X^3 = \tanh \mu (y + y_0) \;, \;\;
  -\arg(w) = \varphi = \varphi_0 \;.\label{sol}
\eeq
%with $w=0$. 

%\end{appendix}

%%%%%%%%%%%%%%%%%%%%%%%%%%%%%%%%%%%%%%%%%%%%%%%%%%%%%%%%%%%%%%%
%\begin{thebibliography}{99}
%
%\bibitem{Hi}
%N.~Hitchin,
%Math. Proc. Cambridge Phil. Soc. {\bf 85}, 465.
%
%\bibitem{NOY}
%T.~Nakatsu, K.~Ohta and T.~Yokono, 
%Phys. Rev. {\bf D58} (1998) 26003, 
%{\tt hep-th/9712029}. 
%
%
%\end{thebibliography}
%
%\end{document}
%%%%%%%%%%%%%%%%%%%%%%%%%%%%%%%%%%%%%%%%%%%%%%%%%%
%
% HSF notation - BPS eq
%
%%%%%%%%%%%%%%%%%%%%%%%%%%%%%%%%%%%%%%%%%%%%%%%%%%
%\include{app-hss1}
%%%%%%%%%%%%%%%%%%%%%%%%%%%%%%%%%%%%%%%%%%%%%%%%%%
% HSF Appendix
%%%%%%%%%%%%%%%%%%%%%%%%%%%%%%%%%%%%%%%%%%%%%%%%%%
% 
%
% Last Modified 8/17/2002 by M.Arai app-hss.tex
%
%               8/24/2002 by M.Arai app-hss1.tex
%
%
%%%%%%%%%%%%%%%%%%%%%%%%%%%%%%%%%%%%%%%%%%%%%%%%%%
\section{The HSF}
\subsection{Notation in the HSF} \label{sc:nothss}
In this Appendix, we summarize our conventions in  
 section \ref{sc:HSF},
 which are mostly the same as those of 
 Refs.~\cite{Ivanov, Ivanov2} and \cite{WB}. 
Here we give only conventions related to the HSF.

Harmonic superspace is defined as 
 $(x^\mu,~\theta_i,~{\bar{\theta}}^i,u_i^{\pm})$ which is called
 the central basis.
The $u_i^{\pm}$ is called the harmonic variable which parameterizes
 the coset $SU(2)_R/U(1)_r\sim S^2$.
The superfield in the HSF is not defined in the central basis 
 but in the subspace which is called the analytic subspace 
 $\{\zeta_A,u_i^{\pm}|x_A^\mu=x^\mu 
  - 2 i \theta^{(i}\sigma^\mu {\bar{\theta}}^{j)}u_{(i}^+u_{j)}^-,
  ~\theta^+ = \theta^i u_i^+,
  ~{\bar{\theta}}^+ = {\bar{\theta}}^i u_i^{+},
  ~u_i^{\pm}\}$,
  where parentheses for indicies $i,j$ mean symmetrization, for instance, 
 $ u_{(i}^+u_{j)}^-=(u_i^+ u_j^- + u_j^+ u_i^-)/2 $.
Hypermultiplet and vector multiplet superfields are defined as the
 function in the analytic subspace as $q^+(\zeta_A,u)$ and
 $V^{++}(\zeta_A,u)$, respectively, which are called 
the analytic superfields.  

To describe the real action in terms of the analytic superfield,
 the star conjugation must be introduced in addition to the
 usual complex conjugation.
The complex conjugation rules for the cofficients in the harmonic
 expansions $f^{i_1\cdots i_n}$ (see (\ref{eq:harmonicexp})), 
 the Grassmann variable 
 $\theta_{i\alpha}$ and the harmonic variable $u_i^{\pm}$
 are defined as
%%%%%%%%%%%%%%%%%%%%%%%%%%%%%%%%
\begin{eqnarray}
 \overline{f^{i_1\cdots i_n}} 
  &\equiv & {\bar{f}}_{i_1\cdots i_n},
\qquad %~~
\overline{f_{i_1\cdots i_n}} 
    = (-1)^n {\bar{f}}^{i_1\cdots i_n}, \\
%   =  - {\bar{f}}^{i_1\cdots i_n}, \\
 \overline{\theta_{i\alpha}} & = &  {\bar{\theta}}_{\dot{\alpha}}^{i}, 
\qquad %~~
\overline{\theta_{\alpha}^i} = -{\bar{\theta}}_{\dot{\alpha}i},\\
\qquad %~~
 \overline{u^{+i}} & = & u_i^-,
\qquad %~~
\overline{u_i^+}=-u^{-i},
\end{eqnarray}
%%%%%%%%%%%%%%%%%%%%%%%%%%%%%%%%
respectively.
The star conjugation rules are defined as
%%%%%%%%%%%%%%%%%%%%%%%%%%%%%%%%
\begin{eqnarray}
 &(f^{i_1\cdots i_n})^* = f^{i_1\cdots i_n},& \\
 &(\theta_{\alpha}^i)^* = \theta_{\alpha}^i,& \\
 &(u^{+i})^{*} = u^{-i},~~(u_i^{+})^{*} = u_i^{-},~~
 (u^{-i})^{*} = -u^{+i},~~(u_i^{-})^{*} = - u_i^{+},&\\
 &(u_i^{\pm})^{**} = -u_i^{\pm}.&
\end{eqnarray}
%%%%%%%%%%%%%%%%%%%%%%%%%%%%%%%%
Note that the star conjugate acts only on the quantity having $U(1)_r$ charge.
We write the combination of the complex and the star conjugation
 as
%%%%%%%%%%%%%%%%%%%%%%%%%%%%%%%%
\begin{eqnarray}
 (\overline{q^+(\zeta_A,u)})^* \equiv \widetilde{q^+}(\zeta_A,u).
\end{eqnarray} 
%%%%%%%%%%%%%%%%%%%%%%%%%%%%%%%%
The combined conjugation rules are defined by
%%%%%%%%%%%%%%%%%%%%%%%%%%%%%%%%
\begin{eqnarray}
&\widetilde{f^{i_1\cdots i_n}}
  =\overline{f^{i_1\cdots i_n}} 
  \equiv  {\bar{f}}_{i_1\cdots i_n}, & \\
& {\widetilde{\theta^+}} = {\bar{\theta}}^+,~~ 
  ~~{\widetilde{\theta^-}} = {\bar{\theta}}^-,
  ~~\widetilde{{\bar{\theta}}^+} = -\theta^+,
  ~~\widetilde{{\bar{\theta}}^-} = -\theta^-, & \\
& (\widetilde{u_i^{\pm}}) = u^{\pm i},
  ~~(\widetilde{u^{\pm i}})=-u_i^{\pm}. &
\end{eqnarray}
%%%%%%%%%%%%%%%%%%%%%%%%%%%%%%%%%%

The simple example of the real action is the free massless action 
 of the Fayet-Sohnius hypermultiplet;
 $ S = -\int d\zeta_A^{(-4)} du~{\widetilde{\phi^+}}D^{++}\phi^+$
 where $D^{++}$ is defined by (\ref{covdhss}) with $Z=0$.
The action is real in the sense of ordinary complex conjugation
 $\bar{S}=S$.
This property follows from the fact that 
 ${\widetilde{\widetilde{q^+}}}=-q^+$.

%%%%%%%%%%%%%%%%%%%%%%%%%%%%%%%%%%
%
% The massive $T^*CP^1$ model
%
%%%%%%%%%%%%%%%%%%%%%%%%%%%%%%%%%%
\subsection{The massive $T^*{\bf C}P^1$ model} \label{sc:TSCP1HSS}
%%%%%%%%%%%%%%%%%%%%%%%%%%%%%%%%%%
In this Appendix, we show how to arrive at the on-shell
 action, and that it is described 
 by on-shell bosonic fields $f_a^i(x_A)$, and their
 superpartners $\psi_a(x_A)$ and $\bar{\varphi}_a(x_A)$. 

We first derive the equations of motion for components.
The Grassmann expansions for analytic superfields are given by
%%%%%%%%%%%%%%%%%%%%%%%%%%%%%
\begin{eqnarray}
 q_a^+(\zeta_A,u) &=& F_a^+ + \sqrt{2} \theta^+ \psi_a 
                      + \sqrt{2} {\bar{\theta}}^+ {\bar{\varphi}}_a 
                      + i\theta^+ \sigma^\mu {\bar{\theta}}^+ A_{a\mu}^-
                      + \theta^+ \theta^+ M_a^- 
                      + {\bar{\theta}}^+{\bar{\theta}}^+ N_a^- 
                        \nonumber \\
                  & & + \sqrt{2} \theta^+\theta^+{\bar{\theta}}^+ 
                        {\bar{\chi}}_a^{--}
                      + \sqrt{2} {\bar{\theta}}^+ {\bar{\theta}}^+
                        \theta^+ \xi_a^{--}
                      + \theta^+ \theta^+ {\bar{\theta}}^+ {\bar{\theta}}^+
                        D_a^{---},
                      \label{eq:exp1} \\
 V_{\rm WZ}^{++}(\zeta_A,u) &=&  \theta^+ \theta^+ {\bar{M}}_v 
                      +{\bar{\theta}}^+ {\bar{\theta}}^+ M_v
                      -2 i \theta^+ \sigma^\mu  {\bar{\theta}}^+ V_\mu
                       \nonumber \\
                   & &  + \sqrt{2} \theta^+\theta^+{\bar{\theta}}^+ 
                        {\bar{\lambda}}^i u_i^-
                      + \sqrt{2} {\bar{\theta}}^+ {\bar{\theta}}^+
                        \theta^+ \lambda^i u_i^{-}
                      +\theta^+ \theta^+ {\bar{\theta}}^+ {\bar{\theta}}^+
                      D_v^{(ij)}u_{(i}^- u_{j)}^-, 
                      \label{eq:exp2}
\end{eqnarray}
%%%%%%%%%%%%%%%%%%%%%%%%%%%%%
where $a=1,2$ is flavor index. 
Note that each component 
 in the hypermultiplet analytic superfield (\ref{eq:exp1})
 includes infinite series expanded by the
 harmonic variable (harmonic expansions), for instance,
%%%%%%%%%%%%%%%%%%%%%%%%%%%%%
\begin{eqnarray}
 F_a^+(x_A,u)
% F_a^+(\zeta_A,u)
 =\displaystyle\sum_{n=0}^{\infty}
   f^{(i_1\cdots i_{n+1}j_1\cdots j_n)}(x_A)
   u_{(i_1}^+ \cdots u_{i_{n+1}}^+u_{j_1}^- \cdots u_{j_n)}^-.
   \label{eq:harmonicexp}
\end{eqnarray}
%%%%%%%%%%%%%%%%%%%%%%%%%%%%%
Thus, the hypermultiplet includes infinite many auxiliary fields in
 addition to physical fields.
Similary, component in the vector multiplet can be expanded in general
 but infinite many auxiliary fields are eliminated by $U(1)$ gauge
 transformation (\ref{gthss3}).
As a result, 
 physical fields $M_v(x_A),~V_\mu(x_A),~\lambda^i(x_A)$
 and auxiliary fields $D_v(x_A)^{(ij)}$ are left as in (\ref{eq:exp2}).
Substituting (\ref{eq:exp1}) and (\ref{eq:exp2}) into 
 (\ref{eq:emhss1})-(\ref{eq:emhss3}), 
 one obtains the equations of motion.
The equation of motion (\ref{eq:emhss1}) reads
%%%%%%%%%%%%%%%%%%%%%%%%%%%%%
\begin{eqnarray}
 \partial^{++} F_1^{+} = 0,~~~\partial^{++} \psi_1 = 0,
 ~~~ \partial^{++} {\bar{\varphi}}_1 &=& 0, \label{appeq11} \\
 \partial^{++} A_{1\mu}^- -2 \partial_\mu^A F_1^+ - 2 V_\mu F_2^+ &=& 0, 
  \label{appeq12} \\
 \partial^{++}M_1^{-} - \frac{\mu}{2} F_1^++{\bar{M}}_v F_2^+ &=& 0, 
  \label{appeq13} \\
 \partial^{++}N_1^{-} + \frac{\mu}{2} F_1^+ +M_v F_2^+ &=& 0, 
  \label{appeq14} \\
 \partial^{++} {\bar{\chi}}_1^{--} - \frac{\mu}{2} {\bar{\varphi}}_1 
  + {\bar{M}}_v {\bar{\varphi}}_2 -i {\bar{\sigma}}^\mu V_\mu {\psi}_2 
  + {\bar{\lambda}}^- F_2^+ -i {\bar{\sigma}}^\mu \partial_\mu^A \psi_1 
  & = & 0,
  \label{appeq15} \\
 \partial^{++} \xi_1^{--} + \frac{\mu}{2} \psi_1 + M_v \psi_2 
  +i \sigma^\mu V_\mu {\bar{\varphi}}_2 + \lambda^- F_2 
  + i\sigma^\mu \partial_\mu^A {\bar{\varphi}}_1 
  & = & 0,
  \label{appeq16} \\
 -\partial_A^\mu A_{1\mu}^- - \frac{\mu}{2} N_1^- + \frac{\mu}{2} M_1^-
  +\partial^{++} D_1^{---} + {\bar{M}}_v N_2^- + M_v M_2^- 
  & & \nonumber \\
  - V^\mu A_{2\mu}^- -{\bar{\lambda}}^- {\bar{\varphi}}_2
  - \lambda^- \psi_2 +D_v^{--} F_2^+ &=& 0,  
  \label{appeq17}
\end{eqnarray}
%%%%%%%%%%%%%%%%%%%%%%%%%%%%%%%%%%
whereas equation (\ref{eq:emhss2}) gives 
%%%%%%%%%%%%%%%%%%%%%%%%%%%%%%%%%%
\begin{eqnarray}
 \partial^{++} F_2^{+} = 0,~~~\partial^{++} \psi_2 = 0,
 ~~~ \partial^{++} {\bar{\varphi}}_2 &=& 0, \label{appeq21} \\
 \partial^{++} A_{2\mu}^- -2 \partial_\mu^A F_2^+ + 2 V_\mu F_1^+ &=& 0, 
  \label{appeq22} \\
 \partial^{++}M_2^{-} - \frac{\mu}{2} F_2^+ - {\bar{M}}_v F_1^+ &=& 0, 
  \label{appeq23} \\
 \partial^{++}N_2^{-} + \frac{\mu}{2} F_2^+ - M_v F_1^+ &=& 0, 
  \label{appeq24} \\
 \partial^{++} {\bar{\chi}}_2^{--} - \frac{\mu}{2} {\bar{\varphi}}_2 
  - {\bar{M}}_v {\bar{\varphi}}_1 + i {\bar{\sigma}}^\mu V_\mu {\psi}_1 
  - {\bar{\lambda}}^- F_1^+ - i {\bar{\sigma}}^\mu \partial_\mu^A \psi_2 
  & = & 0,
  \label{appeq25} \\
 \partial^{++} \xi_2^{--} + \frac{\mu}{2} \psi_2 - M_v \psi_1 
  - i \sigma^\mu V_\mu {\bar{\varphi}}_1 - \lambda^- F_1^+ 
  + i\sigma^\mu \partial_\mu^A {\bar{\varphi}}_2 
  & = & 0,
  \label{appeq26} \\
 -\partial_A^\mu A_{2\mu}^- - \frac{\mu}{2} N_2^- + \frac{\mu}{2} M_2^-
  +\partial^{++} D_2^{---} - {\bar{M}}_v N_1^- - M_v M_1^- 
  & & \nonumber \\
  + V^\mu A_{1\mu}^- +{\bar{\lambda}}^- {\bar{\varphi}}_1
  + \lambda^- \psi_1 - D_v^{--} F_1^+ &=& 0.  
%  - \lambda^- \psi_1 - D_v^{--} F_1^+ &=& 0.  
  \label{appeq27}
\end{eqnarray}
As for constraint (\ref{eq:emhss3}), 
 we write down the relevant part only;
%%%%%%%%%%%%%%%%%%%%%%%%%%%%%%%%%%
\begin{eqnarray}
{\widetilde{F_1^+}} F_2^+ -  {\widetilde{F_2^+}} F_1^+ + \xi^{++} &=& 0, 
 \label{appeq31} \\
{\widetilde{F_1^+}} \psi_2 -F_2^+ \varphi_1 - {\widetilde{F_2^+}} \psi_1
 +F_1^+ {\varphi}_2 &=& 0.  
 \label{eqapp32}
%{\widetilde{F_1^+}} \bar{\varphi}_2 +F_2^+ \bar{\psi}_1 
% - {\widetilde{F_2^+}} \bar{\varphi}_1
% -F_1^+ \bar{\psi}_2 &=& 0,  
% \label{appeq33}
\end{eqnarray}
%%%%%%%%%%%%%%%%%%%%%%%%%%%%%%%%%%
Equations of motion (\ref{appeq11})-(\ref{appeq14}) 
 and (\ref{appeq21})-(\ref{appeq24}) are purely kinematical.
They eliminate the infinite set of auxiliary fields in the 
 harmonic expansions.
The solutions are given as (\ref{solhss1})-(\ref{solhss8}). 
Substituting the component expansion (\ref{eq:exp1}) and (\ref{eq:exp2})
 and on-shell condition  (\ref{solhss1})-(\ref{solhss8})
 into the action (\ref{eq:EHhss}),
 and integrating the Grassmann and the harmonic variable,
 we obtain 
%%%%%%%%%%%%%%%%%%%%%%%%%%%%%%%%%%
\begin{eqnarray}
 S &=& \displaystyle\int d^4 x_A 
       \Bigg\{-\partial_\mu^A f_a^i \partial^\mu_A \bar{f}_{ai}
              -\frac{\mu^2}{4}(f_a^i{\bar{f}}_{ai})^2 
       +\frac{\mid (\mu/2) f_a^i \epsilon_{ab} \bar{f}_{bi}
              -\bar{\psi}_a \epsilon_{ab}\bar{\varphi}_b \mid^2}
                    {f_a^i{\bar{f}}_{ai}}
       \nonumber \\
   & & -i\bar{\psi}_a \bar{\sigma}^\mu \partial_\mu^A \psi_a
       -i\varphi_a \sigma^\mu \partial_\mu^A \bar{\varphi}_a
       -\frac{\mu}{2}(\psi_a \varphi_a  + \bar{\psi}_a \bar{\varphi}_a) 
       \nonumber \\
   & & +\frac{(f_a^i \partial_A^\mu \epsilon_{ab} \bar{f}_{bi}
       -\partial_A^\mu f_a^i \epsilon_{ab} \bar{f}_{bi}
       -i\bar{\psi}_a \bar{\sigma}^\mu \epsilon_{ab} \psi_b
       -i\varphi_a \sigma^\mu \epsilon_{ab} \bar{\varphi}_b)^2}
                    {4f_a^i{\bar{f}}_{ai}} \nonumber \\
   & & 
%+
%-\frac{\lambda^i}{2}\epsilon_{ab}(\varphi_a f_{bi}-\psi_a f_{bi})
%       %-
%+\frac{\bar{\lambda}^i}{2}\epsilon_{ab} (\bar{\psi}_a f_{bi}
%              -\bar{\varphi}_{a}\bar{f}_{bi})
-\frac{\lambda^i}{2}\epsilon_{ab}(\varphi_a f_{bi}-\psi_a {\bar{f}}_{bi})
    +\frac{\bar{\lambda}^i}{2}\epsilon_{ab} (\bar{\psi}_a f_{bi}
                                  +\bar{\varphi}_{a}\bar{f}_{bi})
       -\frac{1}{3} 
       D_{v(ij)}(\epsilon_{ab}\bar{f}^{(i}_a f^{j)}_b+\xi^{(ij)}){\Bigg \}},
        \label{fullachss}
\end{eqnarray}
%%%%%%%%%%%%%%%%%%%%%%%%%%%%%%%%%%
where flavor indices are summed 
 and $\epsilon_{12}=-1,~\epsilon_{21}=1$.
%where $S_{\rm boson}$ is given by (\ref{bosonachss}).
The last line stands for constraints for bosons and fermions.
It is found that full component action is described by the
 hypermultiplet components $f_a^i$ and $\psi_a,~{\bar{\varphi}}_a$.
Note that they are still subject to the constraint (\ref{consthss})
 and $\epsilon_{ab}(\varphi_a f_{bi}-\psi_a {\bar{f}}_{bi})=0$.
In the end of subsection \ref{BPShss}, 
we briefly review how to solve the constraint
 (\ref{consthss}), and rewrite the bosonic part (\ref{bosonachss}) 
of the action 
 (\ref{fullachss}) by independent variables as (\ref{nlsm_lag}).
It is found that the target metric in field space of 
the action (\ref{nlsm_lag})
 coincides with the Eguchi-Hanson metric \cite{CF,ivanov-EH,valent}.
%%%%%%%%%%%%%%%%%%%%%%%%%%%%%%%%%%%%%
%
% BPS equations
%
%%%%%%%%%%%%%%%%%%%%%%%%%%%%%%%%%%%%% 
\subsection{BPS equations in the HSF} \label{ap:BPSeqHSS}
In this Appendix, it is shown that the BPS equations are given as 
 (\ref{eq:BPSeqhss1})-(\ref{eq:BPSeqhss4}).
As mentioned in subsection \ref{BPShss},
 we are interested in the on-shell fermions
 $\psi_a(x_A)$ and ${\bar{\varphi}}_a(x_A)$.
Thus, in the following, we first derive the SUSY transformations 
 for these fermions.
They can be read off
 from the SUSY transformations for the analytic superfields $q_a^+$.
The SUSY transformations for the coordinates on the analytic basis
 are defined as 
%%%%%%%%%%%%%%%%%%%%%%%%%%%%%%%%%%%
\begin{eqnarray}
 \delta_S x_A^\mu  &=& -2 i (\epsilon^i \sigma^\mu {\bar{\theta}}^+ 
                            + \theta^+ \sigma^\mu {\bar{\epsilon}}^i)u_i^- ,
% \delta_S x_a^\mu  &=& -2 i (\epsilon^i \sigma^\mu {\bar{\theta}}^+ 
%                            + \theta^+ \sigma^\mu {\bar{\theta}}^i)u_i^- ,
                            \nonumber \\
 \delta_S \theta^+ &=& \epsilon^i u_i^+,
 ~~\delta_S {\bar{\theta}}^+ = {\bar{\epsilon}}^iu_i^+,~~
 \delta_S u_i^{\pm} = 0, \label{SUSYtrans} \\
 \delta_S x_5 &=& 2 i (\epsilon^i\theta^+-{\bar{\epsilon}}^i 
  {\bar{\theta}}^+)u_i^-, \nonumber \\
 \delta_S x_6 &=& 2 (\epsilon^i \theta^+ 
                 + \bar{\epsilon}^i \bar{\theta}^+)u_i^-,
 %.
 \nonumber  
\end{eqnarray}
%%%%%%%%%%%%%%%%%%%%%%%%%%%%%%%%%%%
where the translations in $x_5$ and $x_6$ 
can be used to generate 
the central charge $Z$ 
by the substitution 
$-i(\partial_5+i \partial_6) \rightarrow \mu/2$. 
In general, $\mu$ is complex but we take it to be real
 as mentioned in section \ref{sc:MTSCP1HSS}.
To derive the SUSY transformations which preserve the Wess-Zumino
 gauge, we calculate the sum of the SUSY transformation and the 
compensating gauge transformation for the
 vector multiplet in the Wess-Zumino gauge, 
%%%%%%%%%%%%%%%%%%%%%%%%%%%%
\begin{eqnarray}
 \hat{\delta}V_{\rm WZ}^{++}=(\delta_S+\delta_G)V_{\rm WZ}^{++}
 =\delta_SV_{\rm WZ}^{++}+D^{++}\lambda, \label{eq:WZhss} 
\end{eqnarray}
%%%%%%%%%%%%%%%%%%%%%%%%%%%%
where $\hat{\delta}$ denotes a sum of SUSY transformation $\delta_S$ and
 gauge transformation $\delta_G$. 
The gauge parameter
 $\lambda(\zeta_A,u)$ is expanded in the component fields as
%%%%%%%%%%%%%%%%%%%%%%%%%%%%
\begin{eqnarray}
 \lambda(\zeta_A,u) &=& F_\lambda+\sqrt{2}\theta^+\psi_\lambda^-
                        +\sqrt{2}{\bar{\theta}}^+{\bar{\psi}}_\lambda^-
                        +\theta^+\theta^+ {\bar{M}}_\lambda^{--}
                        +{\bar{\theta}}^+{\bar{\theta}}^+ M_\lambda^{--} 
                        \nonumber \\
                    & &+i\theta^+\sigma^\mu {\bar{\theta}}^+A_{\lambda \mu}^{--}
                        +\sqrt{2}{\bar{\theta}}^+{\bar{\theta}}^+\theta^+
                         \xi_\lambda^{---}
                        +\sqrt{2}\theta^+\theta^+{\bar{\theta}}^+
                         {\bar{\xi}}_\lambda^{---}
                        +\theta^+\theta^+{\bar{\theta}}^+{\bar{\theta}}^+ 
                         D_\lambda^{(-4)}.\nonumber \\
                         \label{eq:lambda}
\end{eqnarray}
%%%%%%%%%%%%%%%%%%%%%%%%%%%%
Eq.~(\ref{eq:WZhss}) is defined as the new SUSY transformation.
Substituting (\ref{eq:lambda}) into (\ref{eq:WZhss}) 
 and using (\ref{SUSYtrans}) with the understanding 
 $-i\partial_5\rightarrow \mu/2$
 lead to explicit form: 
% of the sum of the SUSY
% transformation and gauge transformation;
%%%%%%%%%%%%%%%%%%%%%%%%%%%%
\begin{eqnarray}
\hat{\delta}V_{\rm WZ}^{++}
&=& \hat{\delta} F_v^{++}+ 
\sqrt{2}\theta^+ \hat{\delta} \psi_v^+
                      +
\sqrt{2}{\bar{\theta}}^+ \hat{\delta} {\bar{\psi}}_v^+
                      +\theta^+\theta^+\hat{\delta} {\bar{M}}_v  
                      +{\bar{\theta}}^+{\bar{\theta}}^+ \hat{\delta}M_v
                      +i\theta^+\sigma^\mu {\bar{\theta}}^+\hat{\delta} V_\mu 
                     \nonumber \\
                 & &+\sqrt{2}{\bar{\theta}}^+{\bar{\theta}}^+\theta^+
                      \hat{\delta} \lambda^-
                      +\sqrt{2}\theta^+\theta^+{\bar{\theta}}^+
                       \hat{\delta}{\bar{\lambda}}^-
                      +\theta^+\theta^+{\bar{\theta}}^+{\bar{\theta}}^+
                      \hat{\delta} D_v^{--}
\end{eqnarray}
%%%%%%%%%%%%%%%%%%%%%%%%%%%%
where
%%%%%%%%%%%%%%%%%%%%%%%%%%%%
\begin{eqnarray}
 \hat{\delta} F_v^{++}(x_A,u) &=& 0 + (\partial^{++}F_\lambda)(x_A,u), \\
 \hat{\delta} \psi_v^+(x_A,u) &=& \sqrt{2}\epsilon^i u_i^+ {\bar{M}}_v(x_A) 
                       -\sqrt{2} i \sigma^\mu {\bar{\epsilon}}^i u_i^+
                       V_\mu(x_A) + (\partial^{++} \psi_\lambda^-)(x_A,u), \\
 \hat{\delta} M_v(x_A,u)      
                 &=& \frac{1}{\sqrt{2}}
                     \epsilon^i \lambda_i(x_A) 
                       +\sqrt{2}\epsilon^{(i}\lambda^{j)}(x_A)
                        u_{(i}^+ u_{j)}^- 
                       +(\partial^{++} M_\lambda^{--})(x_A,u), \\
 \hat{\delta} V_\mu(x_A,u)
                 &=& -\frac{i}{2\sqrt{2}}
                       ({\bar{\epsilon}}^i{\bar{\sigma}}_\mu \lambda_i
                       -\epsilon^i\sigma_\mu{\bar{\lambda}}_i)(x_A)
                       -\frac{i}{\sqrt{2}}
                       ({\bar{\epsilon}}^{(i}{\bar{\sigma}_\mu}\lambda^{j)}
                       -\epsilon^{(i}\sigma_\mu{\bar{\lambda}}^{j)})(x_A)
                       u_{(i}^+u_{j)}^- \nonumber \\
                 & & -\frac{1}{2}(\partial^{++}A_{\lambda \mu}^{--}
                 -2\partial_\mu^A F_\lambda)
                           (x_A,u), \\
 \hat{\delta} \lambda^-(x_A,u)
                 &=& -\sqrt{2}i(\sigma^\mu {\bar{\epsilon}}^i u_i^-)
                 \partial_\mu^A M_v(x_A) -\sqrt{2}\sigma^\nu
                 {\bar{\sigma}}^\mu \epsilon^i u_i^- \partial_\mu^A V_\nu(x_A) 
                 +\sqrt{2}\epsilon^{(i}D_v^{jk)}(x_A)u_{(i}^+u_j^-u_{k)}^-
                 \nonumber \\
                 & &-\frac{2\sqrt{2}}{3}\epsilon_jD_v^{(jk)}(x_A) u_k^- 
                 +(\partial^{++}\xi_\lambda^{---} 
                 +i\sigma^\mu \partial_\mu^A{\bar{\psi}}_\lambda^-)(x_A,u),
                  \\
 \hat{\delta} D_v^{--}(x_A,u)
                 &=& \sqrt{2}i(\epsilon^{(i}\sigma^\mu \partial_\mu^A 
                 {\bar{\lambda}}^{j)}(x_A)
                 -{\bar{\epsilon}}^{(i}{\bar{\sigma}}^\mu
                 \partial_\mu^A \lambda^{j)}(x_A))u_{(i}^-u_{j)}^-
                 +(\partial^{++}D_\lambda^{(-4)}
                 -\partial_A^{\mu} A_{\lambda \mu}^{--})(x_A,u). \nonumber 
                  \\
\end{eqnarray}
%%%%%%%%%%%%%%%%%%%%%%%%%%%%
The last parenthesis for each equation means the contribution from the
 gauge transformation.
%The shifts from the Wess-Zumino
% gauge 
The shifts from the Wess-Zumino gauge are represented by the fields
 with higher harmonic variables which do not appear in Eq.~(\ref{eq:exp2}), 
 and they 
 are pulled back to the original one by taking the gauge parameter
 as
%%%%%%%%%%%%%%%%%%%%%%%%%%%%
\begin{eqnarray}
 F_\lambda(x_A,u)    
           &=& f_\lambda(x_A), \label{gfhss1}\\
 \psi_\lambda^-(x_A,u)  
           &=& -\left(\sqrt{2}\epsilon^i{\bar{M}}_v (x_A)\label{gfhss2}
           -\sqrt{2}i\sigma^\mu{\bar{\epsilon}}^i
           V_\mu(x_A) \right)u_i^-, 
           \label{gfhss3}\\
% M_{\lambda}^{-}(x_A,u)
% M_{\lambda}^{--}(x_A,u)
 M_{\lambda}^{--}(x_A,u)
           &=&-\sqrt{2}{\epsilon}^{(i}\lambda^{j)}(x_A)
           u_{(i}^-u_{j)}^-, \label{gfhss4}
           \\
 A_{\lambda \mu}^{--}(x_A,u)
           &=&-\sqrt{2}i({\bar{\epsilon}}^{(i}
           {\bar{\sigma}}_\mu \lambda^{j)}(x_A)
           -\epsilon^{(i}\sigma_\mu{\bar{\lambda}}^{j)}(x_A))
           u_{(i}^-u_{j)}^-, \label{gfhss5}\\
 \xi_{\lambda}^{---}(x_A,u)   
           &=&-\sqrt{2}\epsilon^{(i}D_v^{jk)}(x_A)
           u_{(i}^-u_j^- u_{k)}^-, 
           \label{gfhss6}\\
 D_\lambda^{(-4)}(x_A,u)
           &=& 0. \label{gfhss7}
\end{eqnarray}
%%%%%%%%%%%%%%%%%%%%%%%%%%%%
At this stage, we can obtain the SUSY transformations $\hat{\delta}$ 
in the Wess-Zumino 
gauge for on-shell fermions, which can be derived from the SUSY 
trandformation for hypermultiplets $\hat{\delta}q_a^+$.
They are defined as
%%%%%%%%%%%%%%%%%%%%%%%%%%%%
\begin{eqnarray}
 \hat{\delta}q_1^+ &=& (\delta_S + \delta_G)q_1^+ 
                 =  \delta_S q_1^+ -\lambda q_2^+, \label{sthss1}\\
%                 =  \delta_S q_1^+ -\lambda q_1^+, \label{sthss1}\\
 \hat{\delta}q_2^+ &=& (\delta_S + \delta_G)q_2^+ 
                 =  \delta_S q_2^+ +\lambda q_1^+. \label{sthss2}
%                 =  \delta_S q_2^+ +\lambda q_2^+. \label{sthss2}
\end{eqnarray}
%%%%%%%%%%%%%%%%%%%%%%%%%%%%
Substituting (\ref{eq:exp1}) and (\ref{gfhss1})-(\ref{gfhss7}) 
 into (\ref{sthss1}) and
 (\ref{sthss2}), 
 the SUSY transformation $\hat{\delta}$ 
for on-shell fermions 
 can be derived as
%%%%%%%%%%%%%%%%%%%%%%%%%%%% 
\begin{eqnarray}
 \hat{\delta}\psi_1(x_A,u) 
                    &=& \sqrt{2}{\epsilon}^i u_i^+ M_1^-
                       -\sqrt{2}i\sigma^\mu {\bar{\epsilon}}^i u_i^-
                        (\partial_\mu^A F_1^++ V_\mu F_2^+) \nonumber \\
                    & &+\frac{i}{\sqrt{2}}\sigma^\mu {\bar{\epsilon}}^i u_i^+
%                    & &+\frac{i}{\sqrt{2}}\sigma^\mu {\bar{\epsilon}}^i u_i^-
                        A_{1\mu}^-
                       -\sqrt{2}\left(\frac{\mu}{2} F_1^+ -{\bar{M}}_v F_2^+
                        \right)\epsilon^i u_i^-,\\
%                        \label{fsv1}, \\
 \hat{\delta}{\bar{\varphi}}_1(x_A,u) 
                    &=& \sqrt{2}{\bar{\epsilon}}^i u_i^+ N_1^-
                       +\sqrt{2}i{\bar{\sigma}}^\mu {\epsilon}^i u_i^-
                        (\partial_\mu^A F_1^++ V_\mu F_2^+) \nonumber \\
%                    & &-\frac{i}{\sqrt{2}}{\bar{\sigma}}^\mu \epsilon^i u_i^-
                    & &-\frac{i}{\sqrt{2}}{\bar{\sigma}}^\mu \epsilon^i u_i^+
                        A_{1\mu}^-
                       +\sqrt{2}\left(\frac{\mu}{2} F_1^+ +M_v F_2^+
                        \right){\bar{\epsilon}}^i u_i^-,\\
%                        \label{fsv2} \\
 \hat{\delta}\psi_2(x_A,u) 
                    &=& \sqrt{2}{\epsilon}^i u_i^+ M_2^-
                       -\sqrt{2}i\sigma^\mu {\bar{\epsilon}}^i u_i^-
                        (\partial_\mu^A F_2^+ - V_\mu F_1^+) \nonumber \\
%                   & &+\frac{i}{\sqrt{2}}\sigma^\mu {\bar{\epsilon}}^i u_i^-
                    & &+\frac{i}{\sqrt{2}}\sigma^\mu {\bar{\epsilon}}^i u_i^+
                        A_{2 \mu}^-
                       -\sqrt{2}\left(\frac{\mu}{2} F_2^+ +{\bar{M}}_v F_1^+
                        \right)\epsilon^i u_i^-,\\
%                        \label{fsv3} \\
 \hat{\delta}{\bar{\varphi}}_2(x_A,u) 
                    &=& \sqrt{2}{\bar{\epsilon}}^i u_i^+ N_2^-
                       +\sqrt{2}i{\bar{\sigma}}^\mu {\epsilon}^i u_i^-
                        (\partial_\mu^A F_2^+ - V_\mu F_1^+) \nonumber \\
%                    & &-\frac{i}{\sqrt{2}}{\bar{\sigma}}^\mu \epsilon^i u_i^-
                    & &-\frac{i}{\sqrt{2}}{\bar{\sigma}}^\mu \epsilon^i u_i^+
                        A_{2\mu}^-
                       +\sqrt{2}\left(\frac{\mu}{2} F_2^+ -M_v F_1^+
                        \right){\bar{\epsilon}}^i u_i^-.
%                        \label{fsv4}
\end{eqnarray}
%%%%%%%%%%%%%%%%%%%%%%%%%%%%
Substituting on-shell condition (\ref{solhss1})-(\ref{solhss8})
 and using $u_i^+ u_j^- -u_i^- u_j^+ = \epsilon_{ij}$, we find
%%%%%%%%%%%%%%%%%%%%%%%%%%%%
\begin{eqnarray}
 \hat{\delta} \psi_1(x_A)
                    &=& -\sqrt{2} \epsilon^i \left({\bar{M}}_v f_{2i}-
                         \frac{\mu}{2} f_{1i}\right)
                        +\sqrt{2} i \sigma^\mu {\bar{\epsilon}}^i
                         (\partial_\mu^A f_{1i} + V_\mu f_{2i}), 
                        \label{fsv1} \\
 \hat{\delta} {\bar{\varphi_1}}(x_A)
                    &=& -\sqrt{2} {\bar{\epsilon}}^i 
                         \left(M_v f_{2i} + \frac{\mu}{2} f_{1i} \right)
                        -\sqrt{2} i {\bar{\sigma}}^\mu \epsilon^i
                         (\partial_\mu^A f_{1i} + V_\mu f_{2i}), 
                         \label{fsv2} \\
 \hat{\delta} \psi_2(x_A)
                    &=& \sqrt{2} \epsilon^i 
                         \left({\bar{M}}_v f_{1i} + \frac{\mu}{2} f_{2i}\right)
                        +\sqrt{2} i \sigma^\mu {\bar{\epsilon}}^i
                         (\partial_\mu^A f_{2i} - V_\mu f_{1i}), 
                         \label{fsv3} \\
  \hat{\delta} {\bar{\varphi_2}}(x_A)
                    &=& \sqrt{2} {\bar{\epsilon}}^i 
                         \left(M_v f_{1i} - \frac{\mu}{2} f_{2i} \right)
                        -\sqrt{2} i {\bar{\sigma}}^\mu \epsilon^i
                         (\partial_\mu^A f_{2i} - V_\mu f_{1i}).
                         \label{fsv4}
\end{eqnarray}
%%%%%%%%%%%%%%%%%%%%%%%%%%%%%%%%%%%%%%
Finally, substituting half SUSY condition (\ref{eq:half-SUSY}), we obtain 
 the BPS equations (\ref{eq:BPSeqhss1})-(\ref{eq:BPSeqhss4}).
%%%%%%%%%%%%%%%%%%%%%%%%%%%%%%%%%%%%%%%%%%%%%%%%%%
%
% KNC
%
%%%%%%%%%%%%%%%%%%%%%%%%%%%%%%%%%%%%%%%%%%%%%%%%%%
%\include{knc}
%%%%%%%%%%%%%%%%%%%%%%%%%%%%%%%%%%%%%%
% KNC part
%%%%%%%%%%%%%%%%%%%%%%%%%%%%%%%%%%%%%%
%
% 8/13/2002 Nitta
%
%\documentstyle[12pt,showkeys]{article}
%\documentstyle[12pt]{article}

%\renewcommand{\baselinestretch}{1.4}
%\textwidth 150mm
%\textheight 220mm

%\makeatletter
%\renewcommand{\theequation}{%
%   \thesection.\arabic{equation}}
% \@addtoreset{equation}{section}
%\makeatother

%\begin{document}
%\topmargin 0pt
%\oddsidemargin 5mm

\section{Mode expansions of boson and fermion} \label{KNC}

In this appendix, we show that 
there exists the correspondence of
bosons and fermions in 
the mode expansions to all orders including massive modes.  
To this end, we extensively use the covariant expansion 
of the Lagrangian in the \kahler normal coordinates (KNC). 
First, we present a brief review of KNC~\cite{HIN}, 
and then we discuss the mode expansion around 
the BPS domain wall background, 
generalising the result of Chibisov and Shifman~\cite{CS}. 

%%%%%%%%%%%%%%%%%%%%%%%%%%%
%%%%%%%%%%%%%%%%%%%%%%%%%%%%%%%%%%%%%%%%%%%%%%%%%%%%%%%%%%

%%%%%
\subsection{Covariant expansion of nonlinear Lagrangian}

%\if0 %%%%%%%%%%
The Lagrangian of ${\cal N}=1$ SUSY nonlinear sigma models 
is given by~\cite{Zu,WB}
\beq
 {\cal L} = \int d^4 \theta \, K(\Phi,\Phi\dagg) 
 + \left[ \int d^2 \theta \, W(\Phi) + {\rm c.c.} \right], 
\eeq
where $\Phi^i(x,\theta,\thb)$ are chiral superfields which consist of 
complex scalar fields $A^i(x)$, Weyl fermions $\psi^i(x)$ 
and auxiliary complex scalar fields $F^i(x)$:
\beq
 \Phi^i(y,\theta) = A^i(y) + \sqrt 2 \theta \psi^i(y) 
  + \theta \theta F^i(y) \,, \hs{5} 
 y^{\mu} = x^{\mu} + i \theta \sig^{\mu} \thb \,,
\eeq
and $K$ is the \kahler potential and 
$W$ is the superpotential.  
%%%%%%%%%%%%%%%%%%%%%%%%%%%%%%%%%%%%%
%SUSY transformation is given by 
%\beq
% && \delta_{\epsilon} A^i = \sqrt 2 \epsilon \psi^i \, ,\non
% && \delta_{\epsilon} \psi^i 
%    = i \sqrt 2 \sig^{\mu} \epb \del_{\mu} A^i 
%    + \sqrt 2 \epsilon F^i \, , \non
% && \delta_{\epsilon} F^i 
%    = i \sqrt 2 \epb \sigb^{\mu} \del_{\mu}\psi^i \, . 
% \label{SUSY-tr.}
%\eeq 
%%%%%%%%%%%%%%%%%%%%%%%%%%%%%%%%%%%%%
Field redefinitions of chiral superfields 
$\Phi^i{}' = f^i(\Phi)$ yield
\beq
 A^i{}' = f^i(A) \, , \hs{5} 
 \psi^i{}' = {\del f^i(A) \over \del A^j} \psi^j,  
 \label{field-redef.}
\eeq 
and redefinitions of $F^i$. 
Therefore, the scalar fields $A^i$ transform as 
coordinates in the target manifold, 
whereas the fermions $\psi^i$ transform as 
a holomorphic tangent vector. 
Equations of motion for $F^i$ 
are  
\beq
 F^i = \1{2} \Gamma^i{}_{jk} \psi^j \psi^k 
     - g^{ij^*} \del_{j^*}W^*, \label{F}
\eeq
where 
%$g_{ij^*}(A,A^*) 
%= \del_i \del_{j^*} K(A,A^*)$ is the \kahler metric 
%and $g^{ij^*}$ is its inverse, 
%and 
$\Gamma^i{}_{jk}(A,A^*) = g^{il^*} g_{jl^*,k^*}(A,A^*)$ is 
the connection. 
We denote differentiation by 
$g_{jl^*,k^*}\equiv \partial g_{jl^*}/\partial A^{k*}$. 
%\fi %%%%%%%%% 

After the elimination of the auxiliary fields, 
 the Lagrangian of SUSY nonlinear sigma models  
 is obtained as
%~\cite{WB}
\beq
% && {\cal L} = 
%    {\cal L}_{\rm kinetic} + {\cal L}_{\rm potential} \\
% && {\cal L}_{\rm kinetic} 
&&{\cal L} 
= - g_{ij^*}
  \del_{\mu} A^i \del^{\mu} A^{*j}
 - i g_{ij^*} 
   \psb^j \sigb^{\mu} D_{\mu} \psi^i 
 + \1{4} R_{ij^*kl^*} \psi^i\psi^k \psb^j\psb^l\, \non
%&& {\cal L}_{\rm potential} 
% = 
&& \hs{8}
- g^{ij^*} D_i W D_{j^*} W^*  
   - \1{2} D_iD_j W \psi^i \psi^j
   - \1{2} D_{i^*}D_{j^*} W^* \psb^i \psb^j \, ,
\eeq
where $D_i$ and $D_{\mu}$ are the covariant derivatives 
on the target space and their pull-back to the space-time, 
respectively: 
\beq
&&D_i W = \del_i W (A) \, , \hs{5} 
  D_i D_j W = \del_i \del_j W (A)
  - \Gamma^k{}_{ij}(A,A^*) \del_k W (A) \,, \non
&&D_{\mu} \psi^i = \del_{\mu}\psi^i 
+ \del_{\mu}A^j \con{i}_{jk}(A,A^*) \psi^k \,. 
\eeq
The Lagrangian is invariant under holomorphic 
field redefinitions, 
%(\ref{field-redef.}), 
corresponding to holomorphic coordinate transformations 
in the target manifold.

%%%%%%%%%%%%%%%%%
%\subsection{Covariant Expansion in Terms of Normal Coordinates}
%

We decompose scalar fields into the 
background fields $\ph^i(x)$ and fluctuating fields 
$\pi^i(x)$ around them:
\beq
 A^i(x) = \ph^i(x) + \pi^i(x) \,. 
\eeq
There appear lots of non-covariant terms in 
the expansion of the Lagrangian in terms of $\pi^i$. 
%
%Using a holomorphic coordinate transformation 
% (\ref{field-redef.}),
%we would like to find the expansion with only covariant tensors.
%Such a covariant expansion can be given by 
%KNC $\omega^i$, which are defined by 
%coordinates such that conditions~\cite{HIN}
%\beq
% \hat K,_{j^*1_1 \cdots i_N}(\omega,\omega^*)|_0 = 0
% \hs{5} \mbox{ or } \hs{5} 
% \del_{i_1} \cdots \del_{i_{N-2}} \hat \Gamma^j{}_{i_{N-1}i_N}
%  (\omega,\omega^*)|_0 = 0 \, ,\label{KNC-cond.}
%\eeq
%hold, where a hat indicates quantities in KNC and 
%the index ``$0$'' denotes a value evaluated 
%at the origin of KNC. 
%
To eliminate such non-covariant terms, 
we transform the fluctuation fields $\pi^i$ 
to KNC fields by~\cite{HIN}
\beq
 \omega^i = \sum_{n=1}^{\infty} \1{n!} 
 [g^{ij^*} K,_{j^*i_1 \cdots i_n}(z,z^*)]_{\ph}  
  \pi^{i_1} \cdots \pi^{i_n} \,,
 \label{KNC-def}
\eeq
where the index ``$\ph$'' denotes a value evaluated 
at the background $A^i = \ph^i$. 
Under holomorphic coordinate transformations of fluctuations
$\pi^i \to \pi^i{}' = \pi^i{}'(\pi)$, 
the KNC $\omega^i$ transform as a holomorphic tangent vector, 
as well as the fermions: 
\beq
 \omega^i \to \omega^i{}' 
   = {\del \pi^i{}' \over \del \pi^j} \omega^j \,,\hs{5}
 \psi^i \to \psi^i{}' 
   =  {\del \pi^i{}' \over \del \pi^j} \psi^j \,.
 \label{KNC-tr.}
\eeq
%Comparing this transformation law with (\ref{field-redef.}), 
%we expect that the expansion using the KNC is
%a covariant expansion. 
%

In KNC, the Taylor expansion gives us a covariant expansion,
because all non-covariant terms vanish. 
Then, using (\ref{KNC-tr.})
we have covariant expansion of each term 
in the Lagrangian in general coordinates as follows:
\beq
 && g_{ij^*} (A, A^*) 
 = g_{ij^*}|_{\ph}
 + R_{ij^*kl^*}|_{\ph} \omega^k \omega^{*l}  + O(\omega^3) \, ,\non
 &&g^{ij^*}(A,A^*) = g^{ij^*}|_{\ph} 
+ R^{ij^*}{}_{kl^*}|_{\ph} \omega^k\omega^{*l} + O(\omega^3) \, ,\non
&& D_i W (A) 
 = D_i W|_{\ph} + D_{j} D_i W|_{\ph} \omega^j 
   + \1{2} D_{j_1} D_{j_2} D_i W|_{\ph} 
      \omega^{j_1} \omega^{j_2} 
   + O(\omega^3) \,, \non
&& \del_{\mu} \pi^i (x) 
 = D_{\mu} \omega^i(x)  
 - \1{2} \del_{\mu}\ph^{*j}(x) R^i{}_{k_1j^*k_2}|_{\ph} 
   \omega^{k_1}(x) \omega^{k_2}(x) + O(\omega^3) \,.
 \label{deriv-exp.2}
\eeq
The covariant derivatives are defined by 
\beq
 &&D_{\mu} \omega^i 
 = \del_{\mu} \omega^i 
  + \del_{\mu}\ph^j \Gamma^i{}_{jk}(\ph,\ph^*) \omega^k , \non
 &&D_{\mu} \psi^i 
 = \del_{\mu} \psi^i 
  + \del_{\mu}\ph^j \Gamma^i{}_{jk}(\ph,\ph^*)\psi^k  \,.
 \label{cov.}
\eeq 
Using the expansion (\ref{deriv-exp.2}),
the Lagrangian can be expanded in terms of KNC as
\beq
 && {\cal L} 
 = {\cal L}^{(0)} + {\cal L}^{(1)} + {\cal L}^{(2)} + O(3) \, ,
\eeq
in which each order of the expansion can be calculated as
\beq
 &&{\cal L}^{(0)} 
 = - g_{ij^*}\del_{\mu}\ph^i \del^{\mu}\ph^{*j}
   - g^{ij^*} D_i W D_{j^*} W^*, \non
 &&
 {\cal L}^{(1)} 
 = - g_{ij^*}( D_{\mu} \omega^i \del^{\mu}\ph^{*j}
            + \del^{\mu}\ph^i D_{\mu} \omega^{*j} ) \non
 && \hs{12}
   -g^{ij^*} [(D_k D_i W)D_{j^*}W^* \omega^k
   + D_i W (D_{l^*}D_{j^*}W) \omega^{*l}], \non  
 &&{\cal L}^{(2)} 
 = - g_{ij^*} D_{\mu} \omega^i D^{\mu} \omega^{*j} \non
 && \hs{12} 
  - R_{ij^*kl^*}
   \left(\omega^k\omega^{*l} \del_{\mu} \ph^i \del^{\mu} \ph^{*j} 
  - \1{2} \omega^i\omega^k \del_{\mu} \ph^{*j} \del^{\mu} \ph^{*l} 
  - \1{2} \omega^{*j}\omega^{*l} \del_{\mu} \ph^i \del^{\mu}
  \ph^k\right) \non 
 && \hs{12} 
 + [g^{mj^*} g^{in^*} R_{mn^*k_1l_1^*}  D_i W D_{j^*} W^* 
    - g^{ij^*}  (D_{k_1} D_i W)(D_{l_1^*} D_{j^*} W^*) ]
   \omega^{k_1} \omega^{*l_1}  \non
 && \hs{12} 
  - \1{2} g^{ij^*} [ (D_{k_1}D_{k_2} D_i W) D_{j^*} W^*
     \omega^{k_1}\omega^{k_2}
    + D_i W (D_{l_1^*}D_{l_2^*}D_{j^*} W^*) 
     \omega^{*l_1}\omega^{*l_2} ] \non
 && \hs{12}
 -  i g_{ij^*} \psb^j \sigb^{\mu} D_{\mu} \psi^i 
 - \1{2} D_iD_j W \psi^i \psi^j 
 - \1{2} D_{i^*}D_{j^*} W^* \psb^i \psb^j \,.
\eeq
All tensors and the connection are 
evaluated at the background fields $\ph$, 
but we use the same notation with the evaluation at 
general fields $A$ unless there is confusion. 
%
%Eq.~(\ref{cov.}) is the direct consequence of the fact 
%that bosons and fermions are on the same basis in 
%our coordinates (\ref{KNC-tr.}). 
%

%%%%
%%%%%%%%%%%%%%%%%%%%%%%%%%%%%%%%%%%%%%%%%%%%%%%%%%%%%%%%
%\section{BPS Domain Walls}
%

\subsection{Expansion around a BPS Domain Wall Background}
We consider (a parallel configuration of) 
BPS domain walls as a background $\ph^i$.
Without loss of generality, we can take 
the spatial direction perpendicular to the walls as $y$. 
The BPS equation for domain walls in the covariant form is
\beq
 \ph^i{}' = - g^{ij^*}(\ph,\ph^*) D_{j^*} W^*(\ph^*),
%         = F^i|_{\ph}
 \label{BPS-eq.} 
\eeq
where the prime denotes the differentiation with 
respect to the spatial coordinate $y$. 
%
%and $F^i|_{\ph}$ represents an evaluation of $F^i$ in (\ref{F}) 
%on a classical background.
%
%Evaluating the SUSY transformation (\ref{SUSY-tr.}) of 
%the fermion, 
%$\delta_{\epsilon} \psi^i|_{\ph} 
%= [i \sqrt 2 \sig^{\mu} \del_{\mu} A^i + \sqrt 2 \epsilon F^i]_{\ph} 
%= i \sqrt 2 (\sig^2 \epb - i \epsilon) F^i|_{\ph}$, 
%on the wall background, 
%the SUSY conservation law, 
%$0 = \delta_{\epsilon} \psi^i|_{\ph}$, gives us 
%\beq
% \sig^2 \epb = i \epsilon \, .
%\eeq
%Hence any solutions of the BPS eq.~(\ref{BPS-eq.}) 
%preserve the half of the original SUSY and 
%therefore are called the half BPS states. 
%

On the BPS domain wall background, 
the covariant derivative $D_{\mu}$ in spacetime reduces to
\beq
 D_{\mu} 
 = \delta_{\mu}^2 (\ph^i{}' D_i + \ph^{*i}{}' D_{i^*}) 
 = - \delta_{\mu}^2 
  (g^{ij^*} D_{j^*} W^* D_i + g^{ij^*} D_i W D_{j^*}) \,,
 \label{cov.deriv.on-wall}
\eeq
where we have used the BPS equation~(\ref{BPS-eq.}).
Acting the operator $D_2$ % = \ph^i{}' D_i + \ph^{*i}{}' D_{i^*}$ 
on the BPS equation~(\ref{BPS-eq.}), we obtain
\beq
 \ph^i{}'' = g^{ij^*} g^{lk^*} D_l W (D_{k^*} D_{j^*}W^*) \,,
 \label{deri.v-BPS-eq.}
\eeq
in which we have used 
$D_k D_{j^*} W^* = \del_k \del_{j^*} W^* = 0$ and  
the metric compatibility $D_k g_{ij^*} = 0$.

We now show that ${\cal L}^{(1)}$ around the wall background, given by
\beq
&&{\cal L}^{(1)} 
 = - g_{ij^*}( D_2 \omega^i \ph^{*j}{}'
            + \ph^i{}' D_2 \omega^{*j} ) \non
 && \hs{12}
   -g^{ij^*} [(D_k D_i W)D_{j^*}W^* \omega^k
   + D_i W (D_{l^*}D_{j^*}W) \omega^{*l}] \, ,
\eeq 
is a total derivative. This can be rewritten as
\beq
 && {\cal L}^{(1)} 
 = - \del_2 [g_{ij^*} 
     (\omega^i \ph^{*j}{}' + \ph^i{}' \omega^{*j})] 
   + g_{ij^*} (\omega^i \ph^{*j}{}'' + \ph^i{}'' \omega^{*j}) \non
&& \hs{12}
   -g^{ij^*} [(D_k D_i W)D_{j^*}W^* \omega^k
   + D_i W (D_{l^*}D_{j^*}W) \omega^{*l}] \, . \label{L1}
\eeq
Substituting (\ref{deri.v-BPS-eq.}) into (\ref{L1}), 
we find 
\beq
 {\cal L}^{(1)} 
 = - \del_2 [g_{ij^*} 
     (\omega^i \ph^{*j}{}' + \ph^i{}' \omega^{*j})] \,.
\eeq

We thus obtain the expansion of Lagrangian 
around the BPS domain wall background, given by 
\beq
 && {\cal L} 
 = {\cal L}^{(0)} + \del_2 (\cdots) \non
 && \hs{8} 
  - g_{ij^*} D_{\mu} \omega^i D^{\mu} \omega^{*j} 
  - g^{ij^*}  (D_{k_1} D_i W)(D_{l_1^*} D_{j^*} W^*) 
              \omega^{k_1} \omega^{*l_1}  \non
 && \hs{8} 
  + \1{2} R_{ij^*kl^*}
   \left(
     g^{mj^*} g^{nl^*} D_m W D_n W \omega^i \omega^k 
   + g^{im^*} g^{kn^*} D_{m^*} W^* D_{n^*}W^* 
                       \omega^{*j}\omega^{*l} \right) \non 
 && \hs{8} 
  - \1{2} g^{ij^*} [ 
     (D_{k_1}D_{k_2} D_i W) D_{j^*} W^* \omega^{k_1}\omega^{k_2}
   + D_i W (D_{l_1^*}D_{l_2^*}D_{j^*} W^*) 
      \omega^{*l_1}\omega^{*l_2} ] \non
 && \hs{8}
 -  i g_{ij^*} \psb^j \sigb^{\mu} D_{\mu} \psi^i 
 - \1{2} D_iD_j W \psi^i \psi^j 
 - \1{2} D_{i^*}D_{j^*} W^* \psb^i \psb^j  + O(3) \,, 
 \label{exp.-wall}
\eeq
where we have used the BPS equation (\ref{BPS-eq.}).

%%%%%%%%%%%%%%%%%%%%%%%%%%%%%%%%%%%%%%%%%%%%%%%%%%%%%%%
%\subsection{Linearlized Equations of Motion for Fluctuations}  
%

Next, we discuss the linearized equation of 
motion for fluctuation. 
We denote the coordinates along the wall by 
$x^a = (t,x,z)$. 
The linearized equations of motion 
can be derived from (\ref{exp.-wall}). 
The equations of motion for $\omega^{*i}$ read 
\beq
&& 0 = D_a D^a \omega^i + D_2 D^2 \omega^i 
 - g^{ij^*} g^{lm^*} (D_k D_l W) (D_{m^*} D_{j^*} W^*) \omega^k \non
&& + [g^{ij^*}g^{pm^*}g^{kn^*} R_{pj^*kl^*} D_{m^*}W^* D_{n^*} W^*
 - g^{ij^*} g^{nm^*} D_n W (D_{m^*} D_{j^*} D_{l^*} W^*) ] 
   \omega^{*l} , \hs{10} 
   \label{LEOM-boson}
\eeq
where we have used the equation 
$g_{ij^*} D_{\mu} \omega^i D^{\mu} \omega^{*j} = 
- \omega^{*j} g_{ij^*} D_{\mu}  D^{\mu} \omega^i 
+ \del_{\mu} ( \omega^{*j} g_{ij^*} D^{\mu} \omega^i)$. 
The equations of motion for $\psb^i$ and $\psi^i$ read
\beq
 &&0 = - i \sigb^a D_a \psi^i - i \sigb^2 D_2 \psi^i 
 - g^{ij^*} (D_{j^*} D_{k^*} W^*) \psb^k \, ,\non 
 &&0 = -i \sig^a D_a \psb^i - i \sigb^2 D_2 \psb^i 
 - g^{ji^*} (D_j D_k W) \psi^k
, \label{LEOM-fermion}
\eeq
respectively. 
In the case of flat target space, 
equations $R_{ij^*kl^*} =0$, $D_{\mu} = \del_{\mu}$ 
and $D_i = \del_i$ 
hold, and therefore 
Eqs.~(\ref{LEOM-boson}) and (\ref{LEOM-fermion}) 
reduce to 
the linearized equation of motion \cite{CS} 
generalized to an arbitrary number of components. 

Now let us decompose complex fields into real fields. 
First, set boson fields $\omega^i \equiv \omega_R^i + i \omega_I^i$. 
Then Eq.~(\ref{LEOM-boson}) is decomposed into 
\beq
0 &=& D_a D^a \omega_R^i 
 - [-\delta^i_l (D_2)^2 
  + g^{ij^*} g^{km^*} (D_k D_l W) (D_{m^*} D_{j^*} W^*)\non
&& - g^{ij^*}g^{pm^*}g^{kn^*} R_{pj^*kl^*} D_{m^*}W^* D_{n^*} W^*
 + g^{ij^*} g^{nm^*} D_n W (D_{m^*} D_{j^*} D_{l^*} W^*) ] 
   \omega_R^l  \non
 &\equiv&  D_a D^a \omega_R^i - {\cal A}^i{}_l \omega_R^l \,, \non
0 &=& D_a D^a \omega_I^i 
 - [-\delta^i_l (D_2)^2 
  + g^{ij^*} g^{km^*} (D_k D_l W) (D_{m^*} D_{j^*} W^*)\non
&& + g^{ij^*}g^{pm^*}g^{kn^*} R_{pj^*kl^*} D_{m^*}W^* D_{n^*} W^*
 - g^{ij^*} g^{nm^*} D_n W (D_{m^*} D_{j^*} D_{l^*} W^*) ] 
   \omega_I^l  \label{LEOM-boson2}
 \non
&\equiv&  D_a D^a \omega_I^i - {\cal B}^i{}_l \omega_I^l \, ,
\eeq
where we have defined two matrix-operators 
${\cal A}^i{}_j$ and ${\cal B}^i{}_j$. 
Second, let us decompose the Weyl fermions into sets of real fermions 
as $\psi = \pmatrix{\psi_1 \cr \psi_2} \equiv \psi_R + i \psi_I$. 
Then their complex conjugates are
\beq
 \sig^2 \psb 
 = \pmatrix {-i \psb^{\dot 2} \cr i \psb^{\dot 1}}
 = i \pmatrix {\psb_{\dot 1} \cr \psb_{\dot 2}}
 = i \pmatrix {\psi_1^* \cr \psi_2^* }
 = i (\psi_R - i \psi_I) \,.
\eeq
After some calculation, 
%we obtain
 it is found that Eq.~(\ref{LEOM-fermion}) is decomposed into
%\beq
% 0 &=& - i \hat D \psi_R^i 
% - [ \delta^i_k D_2  
%    + g^{ij^*} (D_{j^*} D_{k^*} W^*) ] \psi_I^k \non
%  &\equiv& - i \hat D \psi_R^i - ({\cal O}_R)^i{}_k \psi_I^k \,,\non
% 0 &=& - i \hat D \psi_I^i 
% - [ - \delta^i_k D_2  
%    + g^{ij^*} (D_{j^*} D_{k^*} W^*) ] \psi_I^k \non
%  &\equiv& - i \hat D \psi_I^i - ({\cal O}_I)^i{}_k \psi_R^k \,,
% \label{LEOM-fermion2}
%\eeq
\beq
 0 &=& - i \hat D \psi_R^i 
 - [ \delta^i_k D_2  
    + g^{ij^*} (D_{j^*} D_{k^*} W^*) ] \psi_I^k 
  \equiv - i \hat D \psi_R^i - ({\cal O}_R)^i{}_k \psi_I^k \,,\non
 0 &=& - i \hat D \psi_I^i 
 - [ - \delta^i_k D_2  
    + g^{ij^*} (D_{j^*} D_{k^*} W^*) ] \psi_I^k 
  \equiv - i \hat D \psi_I^i - ({\cal O}_I)^i{}_k \psi_R^k \,,
 \label{LEOM-fermion2}
\eeq
where we have defined the matrix-operators 
${\cal O}_R$ and ${\cal O}_I$, and $\hat D$ is defined by
\beq
 \hat D_{\alpha}{}^{\beta} 
 = \sig^2{}_{\alpha\dot\beta} \sigb^{a \dot\beta\beta} D_a 
 = - \sig^a{}_{\alpha\dot\beta} \sigb^{2 \dot\beta\beta} D_a \,.
\eeq

%%%%%%%%%%%%%%%%%%%%
\subsection{Mode expansions}
Finally, we show the correspondence of boson and fermion 
in the mode expansion. 
To this end, we show that the remarkable relations between 
the matrix-operators for bosons and fermions, 
defined in Eqs.~(\ref{LEOM-boson2}) and (\ref{LEOM-fermion2}), 
hold:
\beq
({\cal O}_R)^i{}_k ({\cal O}_I)^k{}_j = {\cal B}^i{}_j  \,,\hs{5} 
 ({\cal O}_I)^i{}_k ({\cal O}_R)^k{}_j = {\cal A}^i{}_j \,.
 \label{operator-rel.}
\eeq
%As seen below, these equations play an important role 
%in analyses of SUSY in the effective field theory on the wall.
These can be shown as follows: 
\beq
 &&({\cal O}_R)^i{}_k ({\cal O}_I)^k{}_j 
 = - \delta^i_j (D_2)^2 
  + g^{im^*} g^{kl^*} 
    (D_{k} D_{m} W) (D_{l^*} D_{j^*} W^*) 
% \non && \hs{30} 
  + D_2 [g^{il^*} (D_{l^*} D_{j^*} W^*)] \, . \hs{5}
\eeq
Using Eq.~(\ref{cov.deriv.on-wall}), 
the last term can be rewritten as 
\beq
 && D_2 [g^{il^*} (D_{l^*} D_{j^*} W^*)] \non
 && = 
 - g^{mn^*} g^{il^*} D_{n^*} W^* (D_m D_{l^*} D_{j^*} W^*) 
 - g^{il^*} g^{nm^*} D_n W (D_{m^*} D_{l^*} D_{j^*} W^*)\, .\hs{5}  
\eeq
The term in the parenthesis can be calculated, to give
\beq
 D_m D_{l^*} D_{j^*} W^* 
 = R_{ml^*j^*}{}^{p^*} D_{p^*} W^* + D_{l^*} D_m D_{j^*} W^* 
 =  R_{ml^*j^*}{}^{p^*} D_{p^*} W^* \,,
\eeq
in which the equation 
$D_m D_{j^*} W^* = \del_m \del_{j^*} W^* = 0$ has been used. 
We thus have shown the first equation in (\ref{operator-rel.}). 
The second one can be shown in the same way. 

\if0%%%%%%%%%%%
The operators for the fermions in the expansion in terms of $\pi^i$
are the same as ${\cal O}_R$ and ${\cal O}_I$ 
in the KNC expansion, because the fermions always transform 
as a holomorphic tangent vector as in Eq.~(\ref{field-redef.}); 
on the other hand the operators for bosons in $\pi^i$ 
are completely different from 
${\cal A}^i{}_j$ or ${\cal B}^i{}_j$ in KNC. 
Hence we conclude that the KNC expansion is 
the only expansion such that relations 
(\ref{operator-rel.}) hold. 
\fi%%%%%%%%%%%%%

%%%%%%%%%%%%%%%%%%%%%%%%%%%%%
%\subsection{Mode Expansions}

Using the operator relation (\ref{operator-rel.}), 
we discuss the mode expansion.
Prepare sets of the mode functions $a_n^j (y)$ and $b_n^j (y)$ 
($n=1,2,\cdots$) satisfying 
\beq
 ({\cal O}_R)^i{}_j a_n^j (y) = \kappa_n b_n^i (y) \,, \hs{5} 
 ({\cal O}_I)^i{}_j b_n^j (y) = \lam_n a_n^i (y) \,. 
\eeq
Then these functions satisfy 
\beq
 ({\cal O}_I {\cal O}_R)^i{}_j a_n^j (y) 
   = (m_n)^2 a_n^i(y) \,,\hs{5} 
 ({\cal O}_R {\cal O}_I)^i{}_j b_n^j (y) 
   = (m_n)^2 b_n^i(y) \,,
\eeq
where we have defined $m_n = \sqrt {\kappa_n \lam_n}$. 
Therefore 
$a_n \equiv \{a_n^i \}$ ($b_n \equiv \{b_n^i\}$) is an eigenvector 
of the matrix operator ${\cal O}_I {\cal O}_R$  
(${\cal O}_R {\cal O}_I$) with an eigenvalue $(m_n)^2$. 
Using a complete set of $a_n$ and $b_n$, 
the fluctuating fields can be expanded as
\beq
 &&\omega_R^i = \sum_n \sqrt{\lam_n} a_n^i(y) 
      \omega_R^{(n)}(x^a) \,,\hs{5} 
   \omega_I^i = \sum_n \sqrt{\kappa_n} b_n^i(y) 
      \omega_I^{(n)}(x^a) \,,\non 
 &&\psi_R^i = \sum_n \sqrt{\lam_n} a_n^i(y) 
      \psi_R^{(n)}(x^a) \,,\hs{5} 
   \psi_I^i = \sum_n \sqrt{\kappa_n} b_n^i(y) 
      \psi_I^{(n)}(x^a) \,. 
\eeq

The linearized equations of motion (\ref{LEOM-boson2}) and 
(\ref{LEOM-fermion2}) reduce to
\beq
 D_a D^a \omega_R^{(n)} - (m_n)^2 \omega_R^{(n)}  = 0 \,,&& \hs{5} 
   D_a D^a \omega_I^{(n)} - (m_n)^2 \omega_I^{(n)}  = 0 \,, \non
 -i \hat D \psi_R^{(n)} - m_n \psi_I^{(n)} = 0 \,,&& \hs{5} 
   -i \hat D \psi_I^{(n)} - m_n \psi_R^{(n)} = 0 \,,
\eeq
which are equations of motion in the three dimensional 
effective field theory on the wall.

%\end{appendix}

%%%%%%%%%%%%%%%%%%%%%%%%%%%%%%%%%%%%%%%%%%%%%%%%%%%%%%%%%%%%%%%
%\begin{thebibliography}{99}

%\bibitem{Zu}
%B.~Zumino,
%Phys. Lett. {\bf 87B} (1979) 203.\\  
%L.~Alvarez-Gaum\'{e} and D.~Z.~Freedman, 
%Comm. Math. Phys. {\bf 80} (1981) 443.
%
%%%%% KNC
%\bibitem{AGHKLR}
%L.~Alvarez-Gaum\'{e} and P.~Ginsparg, 
%Comm. Math. Phys. {\bf 102} (1985) 311.\\
%C.~M.~Hull, A.~Karlhede, U.~Lindstr\"om and M.~Ro\v{c}ek, 
%Nucl. Phys. {\bf B266} (1986) 1.
%

%\bibitem{HIN}
%K.~Higashijima and M.~Nitta, 
%Prog. Theor. Phys. {\bf 105} (2001) 243, {\tt hep-th/0006027}.\\
%K.~Higashijima, E.~Itou and M.~Nitta, 
%Prog. Theor. Phys. {\bf 108} (2002) 185, {\tt hep-th/0203081}.

%
%\bibitem{AFM}
%L.~Alvarez-Gaum\'{e}, D.~Z.~Freedman and S.~Mukhi,
%Ann. of Phys. {\bf 134} (1981) 85.
%

%%%%
%\bibitem{NNS}
%M.~Naganuma, M.~Nitta and N.~Sakai, 
%Phys. Rev. {\bf D65} (2002) 045016, 
%{\tt hep-th/0108179}.

%\bibitem{CS} 
%B.~Chibisov and M.~Shifman, Phys. Rev. {\bf D56} (1997) 7990, 
%{\tt hep-th/9706141}.

%%
%\bibitem{WB}
%J.~Wess and J.~Bagger,
%{\em Supersymmetry and Supergravity}, 
%Princeton Univ. Press, Princeton (1992).
%
%\end{thebibliography}
%\end{document}
%%%%%%%%%%%%%%%%%%%%%%%%%%%%%%%%%%%%%%%%%%%%%%%%%%
%
%
% Reference 
%
%
%%%%%%%%%%%%%%%%%%%%%%%%%%%%%%%%%%%%%%%%%%%%%%%%%%

\end{document}